\documentclass{aa}  
\usepackage{graphicx}
\usepackage{txfonts}
\newcommand{\dd}{\mathrm{d}}

\usepackage{xcolor}

\usepackage{ulem}

\usepackage{hyperref}
\usepackage{url}

\usepackage{natbib}
\def\gsim{\lower.5ex\hbox{$\; \buildrel > \over \sim \;$}}
\def\lsim{\lower.5ex\hbox{$\; \buildrel < \over \sim \;$}}
\begin{document}

   \title{Particle acceleration at radiative supernova remnant shocks}

  % \subtitle{I. Overviewing the $\kappa$-mechanism}

   \author{P. Cristofari
          \inst{1}
          }

   \institute{Laboratoire d’étude de l’Univers et des phénomènes eXtrêmes, LUX, Observatoire de Paris, Université PSL, Sorbonne Université, CNRS, 92190 Meudon, France\\
              \email{pierre.cristofari@obspm.fr}
             }

   \date{Received September 15, 1996; accepted March 16, 1997}

% \abstract{}{}{}{}{} 
% 5 {} token are mandatory
 
  \abstract
   {Numerous astrophysical shock waves evolve in an environment where the radiative cooling behind the shock affects the hydrodynamical structure downstream, thereby influencing the potential for particle acceleration via diffusive shock acceleration (DSA). }
   {We study the possibility for DSA to energize particles from the thermal pool and from preexisting cosmic rays at radiative shocks, focusing on the case of supernova remnants (SNRs).}
   {We relied on a semi-analytical description of particle acceleration at collisionless shocks in the test-particle limit, estimating the particle spectrum, maximum energy, and total proton and electron content expected from SNRs throughout the radiative phase.}
   {Our results indicate that DSA at radiative shocks can lead to significant particle acceleration during the first few tens of kiloyears of the radiative phase. Although the associated multiwavelength emission from SNRs in the radiative phase may not be detectable with current observatories in most cases, the  radiative phase is found to lead to substantial deviations from the canonical p$^{-4}$ of the test-particle limit. The hardening and/or steepening is due to an interplay between a growing contribution of the reaccelerated term as the SNR volume expands and the effects of adiabatic and radiative losses on trapped particles as particles are confined for a longer time. The slope of the cumulative proton and electron spectra over the SNR lifetime thus depends on the environment in which the SNR shock propagates, and on the duration of the radiative phase during which DSA can take place. Overall, DSA in the radiative phase can lead to a total electron spectrum steeper than the proton spectrum, both at SNRs from  thermonuclear and core--collapse SNe.  Finally, we comment on the case of young  radiative SNRs (in the first month to a few years after the explosion) for which the denser environments (with mass-loss rates of $\dot{M} \sim 10^{-1} - 1$ M$_{\odot}$/yr) tend to inhibit DSA.}
   {}


   \maketitle
%
%-------------------------------------------------------------------

\section{Introduction}
Astrophysical shock waves are known to be the place of efficient particle acceleration up to the very high-energy (VHE) domain through diffusive shock acceleration (DSA)~\citep{axford1977,krymskii1977,bell1978,blandford1978}. The observation of VHE gamma rays at various shock waves, such as supernova remnants (SNRs)~\citep{HESS_SNRs}, gamma-ray bursts~\citep{piran2004}, large-scale shocks in galaxy clusters~\citep{blasi2007}, relativistic jets in active galactic nuclei (AGNs)~\citep{valtaoja1995}, or pulsar wind nebulae~\citep{bednarek2003} is seen as a direct indication that strong collisionless adiabatic shocks are indeed accelerating particles through DSA. However, the ability of radiative shocks to accelerate particles remains a subject of debate.

In radiative shocks, efficient post-shock cooling significantly alters the hydrodynamical structure, causing the downstream temperature and velocity drop rapidly, while the density rises. These conditions are generally unfavorable for efficient particle acceleration~\citep{raymond1979, binette1985, bertschinger1986, cioffi1988, drake2005}. Several factors contribute to this limitation:

1)  The loss of a large amount of energy to radiation reduces the energy available for acceleration; 
2) the lower temperature weakens the generation of turbulent motions in the post-shock region, in turn leading to weaker magnetic turbulence -- crucial for scattering particles back and forth across the shock; 
3) the rise of the post-shock density tends to  make the plasma collisional, in which suprathermal particles can be thermalized before they enter the acceleration process.
4) the residence time of particles in the thin layer around the shock is reduced, thereby limiting the acceleration process.

Nevertheless, in many cases, cooling is not instantaneous, and on some level, the physical conditions close to the shock discontinuity  may still support DSA. Thus, some radiative shocks might energize particles from the thermal pool or reaccelerate preexisting cosmic rays (CRs). 
Radiative shocks have been observed or predicted in a wide variety of environments, including accretion shocks of young stellar objects~\citep{matsakos2013}, bow shocks of runaway stars~\citep{Carretero-Castrillo2025}, 
perturbed magnetospheres of neutron stars~\citep{beloborodov2023}, or accreting black holes~\citep{okuda2021}.

The detection of novae in the VHE gamma-ray domain is a notable example of radiative shocks efficiently accelerating particles via DSA~\citep{metzger2015,li2017,vurm2018,HESS_RSophiuchi2022,phan2025}.

The case of SNRs is particularly instructive, as DSA is clearly active during the free expansion (FE) and adiabatic phases, but often overlooked during the radiative phase of evolution~\citep{ptuskin2005, reynolds2008, schure2010, caprioli2012, celli2019}. While multiwavelength observations~\citep{vink2012} do not support efficient DSA during the radiative phase, which typically begins 20-40 kyr after the supernova (SN) explosion (for a Type Ia, and $\lesssim 10$ kyr for core-collapse SNe), recent studies suggest that the radiative phase could still contribute significantly. It may lead to hard spectra ($\propto p^{-3}$) of accelerated particles, such as protons, electrons, and nuclei, up to several hundred giga-electronvolts~\citep{zirakashvili2022}, and enhance the brightness from radio to VHE gamma rays~\citep{diesing2024, diesing2025}. A similar scenario arises in very young SNRs, where shocks expand into the dense wind of the progenitor star, which can become radiative and hinder particle acceleration due to high cooling rates~\citep{fang2020, pitik2023}.

In this paper, we examine the radiative phase of SNRs and investigate its potential for particle acceleration. The paper is organized as follows. Section 2 describes the structure of radiative shocks and the associated mechanisms of particle acceleration. Section 3 applies this framework to the case of SNRs. In Section 4, we present and discuss our results. Finally, Section 5 summarizes our conclusions.

\section{Radiative shocks}
\label{sec:rad}
\subsection{Shock structure}
In astrophysical plasmas with temperatures of $T \gtrsim 10^4$ K, hydrogen is sufficiently ionized so that collisional excitations of atoms and ions are primarily driven by electron collisions. At low densities, these excitations are typically followed by radiative decay, resulting in energy loss. The associated radiative cooling function generally takes the form

\begin{equation}
f_{\rm cool} (T) = \frac{\Lambda}{n_e n_{\rm H}}
,\end{equation}
with $n_e$ and $n_{\rm H}$ the density of electrons and hydrogen, respectively, and $\Lambda$ the rate of dissipation of thermal energy per unit volume~\citep{draine2011}.
A decent approximation for $f_{\rm cool}$ was proposed in~\citet{draine2011}: 

\begin{align}
f_{\rm cool}(T) &\approx 1.1 \times 10^{-22} 
\left( \frac{T}{10^6\,\text{K}} \right)^{-0.7} \text{erg cm}^{3} \text{s}^{-1}\notag \\
&\quad \text{for } 10^5\,\text{K} < T < 10^{7.3}\,\text{K} \\
f_{\rm cool}(T) &\approx 2.3 \times 10^{-24} 
\left( \frac{T}{10^6\,\text{K}} \right)^{0.5} \text{erg cm}^{3} \text{s}^{-1}\notag \\
&\quad \text{for } T > 10^{7.3}\,\text{K}
.\end{align}

Another widely used prescription is from~\citep{chevalier1994}:
\begin{align}
f_{\rm cool}(T) &\approx  6.2 \times 10^{-19} T^{-0.6} \text{erg cm}^{3} \text{s}^{-1}\notag \\
&\quad \text{for }10^5 < T < 4.7 \times 10^{7} \text{K}  \\
f_{\rm cool}(T) &\approx  2.5 \times 10^{-27} T^{0.5} \text{erg cm}^{3} \text{s}^{-1}\notag \\
&\quad \text{for } 4.7 \times 10^{7} \text{K}   < T
;\end{align}
this expression is adopted as a reference in the following. The typical timescale associated with cooling, in the case of isobaric cooling, is $\tau_{\rm cool} \sim \frac{3 nk_{\rm B} T }{2 \Lambda}$.

Let us consider the case of a one-dimensional, infinite plane shock. 
The physical quantities (velocity, pressure, density, and temperature)  downstream can be estimated through the Rankine-Hugoniot conditions, obtained by writing the conservation of mass, momentum, and energy. Using index 1 for quantities upstream and 2 for downstream: 
\begin{align}
\frac{u_2}{u_1}= \frac{\rho_1}{\rho_2} =  \frac{(\gamma -1) {\cal M}_1^2 +2 }{(\gamma +1) {\cal M}_1^2}\\
\frac{T_2}{T_1} = \frac{[(\gamma -1){\cal M}_1^2+2][2\gamma {\cal M}_1^2- (\gamma -1)]}{(\gamma + 1)^2 {\cal M}_1^2}
,\end{align}
where $\gamma$ is the adiabatic index, and ${\cal M}_1 = \frac{u_1}{{c_s}_1} = \left(\frac{\rho_1 u_1^2}{\gamma P_1} \right)^{1/2}$ is the upstream Mach number, ${c_s}_1$ being the sound speed upstream. 
For strong shocks where ${\cal M_1} \gg 1$, and assuming $\gamma = 5/3$, the above relations reduce to $u_1/u_2=\rho_2/\rho_1 \rightarrow \frac{\gamma +1}{\gamma -1}= 4$, and $k_{\rm B} T_2 \rightarrow \frac{2 \gamma (\gamma -1)}{(\gamma +1)^2}= 3/16 m u_1^2$.

To take into account the cooling in the mass, momentum, and energy conservation equation, the cooling function, $\Lambda$, needs to be included downstream. 

\begin{align}
&\frac{\dd \rho  u}{\dd x}= 0 \label{eq:syst1} \\
&u \frac{\dd u}{ \dd x} = - \frac{1}{\rho} \frac{\dd P}{ \dd x}  \label{eq:syst2} \\
&\frac{1}{\gamma -1}\frac{\dd P u}{\dd x} = - P \frac{\dd u}{\dd x}-  \Lambda(T, \rho) 
\label{eq:syst3}
\end{align}

\begin{figure}
    \centering
	\includegraphics[scale=0.5]{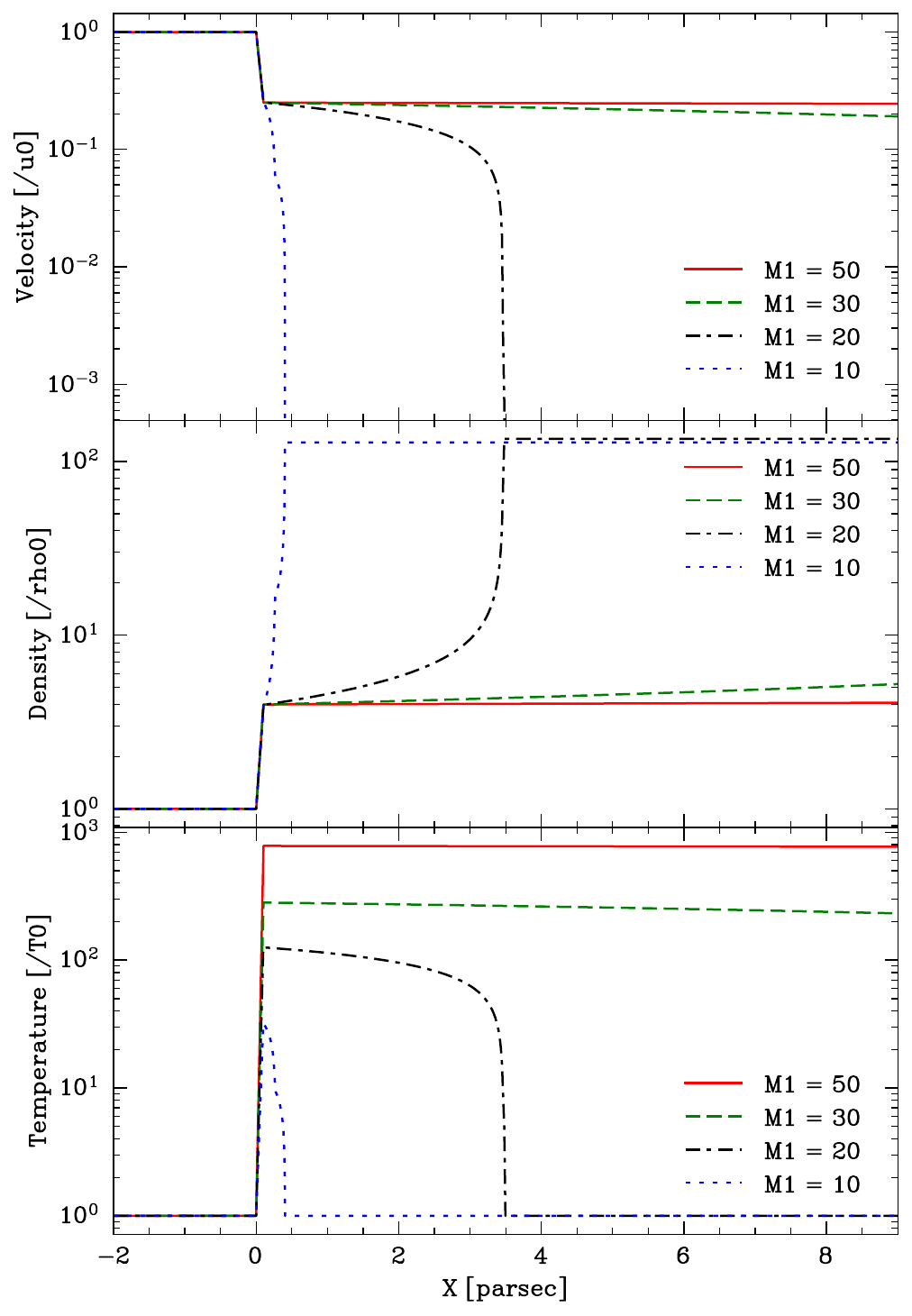}
    \caption{Velocity, density, and temperature of the fluid across the shock, for Mach numbers upstream ${\cal M}_1$=10 (dotted blue), 20 (dash-dotted black), 30 (dashed green), and 50 (solid red).}
    \label{fig:shock_structure}
\end{figure}

The cooling term, $\Lambda \propto f_{\rm cool}$, becomes important when the temperature downstream decreases, i.e., when the shock speed decreases. The effect of cooling on the shock structure is illustrated in Fig.~\ref{fig:shock_structure}, solving Eq.~\eqref{eq:syst1}-\eqref{eq:syst3} with an Eulerian scheme. The integration was started just downstream of the shock front with post-shock values determined from the Rankine--Hugoniot conditions. The solution was then extended downstream until the temperature reaches a prescribed floor value, at which point outflow (zero-gradient) boundary conditions were imposed.

\subsection{Particle acceleration}
\label{sec:pmax}
The spectrum of accelerated particles at the shock can be estimated by solving the transport equation at a one-dimensional infinite plane shock,

\begin{equation}
-u_1 f_0 (p) - \frac{1}{3}(u_1-u_2(p)) p \frac{\partial f_0}{\partial p} + q_0(p)=0
,\end{equation}
where $f_0$ is the spectrum of accelerated particles at the shock, and $q_0= \frac{\eta u_1 n_1}{4 \pi p_{\rm inj}^2} \delta (p-p_{\rm inj})$ the injection term at the shock. At radiative shocks, the cooling induces variations in the fluid velocity downstream, i.e., $u_2$ is a function of the space coordinate (x). Assuming that particles of momentum $p$ can diffuse up to a distance of $x \sim D_2/u_2$, the dependence on the space coordinate can be written as a dependence on momentum. The spectrum at the shock obtained is
\begin{equation}
f_0(p) = \frac{3 r_{\rm sub}}{r_{\rm sub} - 1}\frac{\eta n_1}{4 \pi p_{\rm inj}^3} \exp \left[ \int_{p_{\rm inj}}^p  \frac{3 R_{\rm tot}(p)}{R_{\rm tot}(p)-1} \frac{\text{d}p'}{p'}\right]
\label{eq:f0}
,\end{equation}
where $r_{\rm rub}$ is the compression ratio from upstream to immediately downstream the shock, and $R_{\rm tot}= u_1/u_2(p)$ is the overall compression factor. Several studies have examined in detail the particle spectrum at shocks, accounting for a range of physical effects such as the presence of preexisting CRs \citep{blasi2002, blasi2004}, nonlinear feedback from efficient particle acceleration \citep{amato2005, amato2006}, and modifications due to spatial variations in the velocity profile \citep{petruk2024}. While these effects are certainly important in specific contexts, they are beyond the scope of the present work and are therefore not included in our model.

The maximum momentum of the accelerated particles, $p_{\rm max}$, is not included in Eq.~\eqref{eq:f0} and can be accounted for by introducing an exponential suppression. For DSA to occur, the shock needs to be strong, and the medium “sufficiently” ionized. We assumed that the condition on ionization was met~\citep{sutherland2017}, if only because we were considering shocks expanding in the warm ISM. 
As the shock slows down, the Mach number decreases. We verified that the shock remains supersonic ($\mathcal{M} > 1$), and assumed that DSA can occur at relatively weak shocks with Mach numbers of $\mathcal{M} \sim$ 2-3. Although particle acceleration is typically associated with strong, high-Mach-number shocks, both theoretical and observational studies have shown that even low-Mach-number shocks ($\mathcal{M} \sim$ 2- 3) can accelerate particles under favorable conditions.
 In particular, hybrid and particle-in-cell simulations have demonstrated efficient ion and electron acceleration at quasi-parallel low Mach shocks~\citep{guo2013,caprioli2014,park2015}. This has been further explored in the context of high-beta astrophysical plasmas, such as those found in galaxy clusters, where low Mach shocks are common~\citep{guo2014a,guo2014b}. Cosmological simulations also support the role of such shocks in generating CRs~\citep{wittor2017}. In situ measurements in the solar wind provide additional observational evidence of particle acceleration at low-Mach-number collisionless shocks~\citep{wilson2013}
  
Moreover, for DSA to take place, the shock needs to be collisionless.  A usual criterion to characterize collisionless shocks relies on the idea that the transition from upstream to downstream occurs on a scale smaller than the mean free path (mfp) of particles. 
The mfp downstream can be estimated at the time needed for a projectile to be deflected by $90^{\circ}$, and typically reads (\citep{draine2011}):
\begin{equation}
\text{mfp}= \frac{m^2 v^4}{8 \pi n_2 e^4 \ln \Lambda} \approx 5 \times 10^{17} \left( \frac{T}{10^6 \text{K}}\right) \left( \frac{n_e}{0.01 \text{cm}^{-3}}\right)^{-1}\left( \frac{\Lambda}{25}\right)^{-1}\text{cm}
.\end{equation}

We assumed that particles can be efficiently accelerated as long as their mfp is much larger than their Larmor radius (mfp $\gg r_{\rm L}$), and used this criterion to estimate the maximum energy of accelerated particles. One could argue, however, that the conditions for acceleration may no longer hold in regions where cooling significantly affects the shock structure (where the temperature decreases and the density increases) and where the shock may no longer remain collisionless. In practice, this is not a concern, since the Larmor radius remains smaller than the typical cooling length ($r_{\rm L} \lesssim L_{\rm cool}$).

\begin{figure}
    \centering
	\includegraphics[width=2.9in]{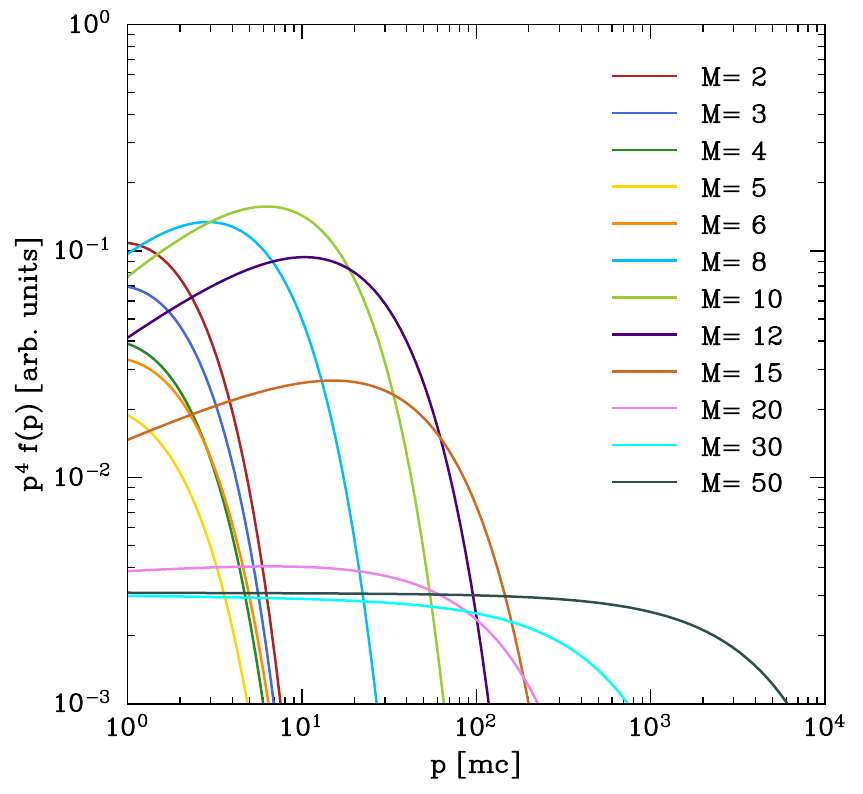}
    \caption{Spectra of accelerated particles at the shocks for various Mach numbers. }
    \label{fig:spectra}
\end{figure}

The typical spectra at the shock are illustrated in Fig.~\ref{fig:spectra}. For low Mach numbers (${\cal M} \sim 2-3$), the maximum momentum of accelerated particles is low enough that the slope  of accelerated particles ($\alpha \rightarrow 3$) is not visible. For large Mach numbers ${\cal M} \sim 50$), $\alpha \rightarrow 4$ is recovered for a strong shock.

\subsection{Limiting effects}
\label{sec:damping}

Ion-neutral damping can be important at SNR shocks propagating through a partially ionized medium, as it can suppress the growth of magnetohydrodynamic waves, and thus limit magnetic field amplification, and influence CR acceleration. As the temperature drops downstream, the gas downstream can recombine more easily, and thus the fraction of neutral hydrogen can increase, favoring ion-neutral damping. Moreover, slower shocks are less efficient at ionizing the upstream medium, allowing for a significant fraction of neutral to pass downstream. Ion-neutral damping has been discussed upstream~\citep{draine1993,drury1996,bykov2000,ptuskin2003} and downstream~\citep{achterberg1986}.

As was discussed in~\citet{drury1996}, ions and neutral are coupled ('oscillating') together provided that the wave frequency remains smaller than the typical ion-neutral collision frequency. If the wave frequency is greater than the collision frequency, however, ions and neutrals can be treated as decoupled. In other words, two regimes (coupled and decoupled) are delimited by an energy, $E_{\rm coup}$, and CRs of energy $E>E_{\rm coup}$ are in the coupled regime. The energy, $E_{\rm damp}$, above which the damping to ion-neutral collisions becomes important was estimated by equating the flux of accelerated particles advected downstream, and the flux of particles leaving the system, due to the damped waves incapable of confining them. When $E_{\rm damp}>E_{\rm coup}$, the ion-neutral damping is considered to be inefficient; this condition can be rewritten as~\citep{padovani2015, padovani2016} 
\begin{equation}
    {\cal R}\approx 1.4 \; 10^{-1} \; \beta^{-1}  \Xi \frac{x^{1.5}}{1-x} \left( \frac{u_{\rm sh}}{10^2 \text{km/s}}\right)^3 \left( \frac{n}{1 \text{cm}^{-3}} \right) \left( \frac{B}{ 3 \mu\text{G}} \right)^{-6} \left(\frac{\xi}{0.01} \right) >1 
  \label{eq:R_1}
,\end{equation}

with $x$ being the ionization fraction, and 
\begin{equation}
\Xi \approx 8 \; 10^{-3}\left( \frac{B}{3 \mu\text{G}}\right)^{4} + 1.4 \; 10^{-6} \Gamma^2 \beta^2 \tilde{\mu} \left(\frac{T}{10^{4}\text{K}} \right)^{0.8} \left( \frac{n}{1\text{cm}^{-3}} \right)^3 x^2
,\end{equation}
where $\Gamma$ is the particle Lorentz factor, $\beta=\Gamma^{-1}(\Gamma^2-1)^{1/2}$, and $\tilde{\mu}=m/m_{p}$ is the mean particle mass in units of proton mass. Typically, $\tilde{\mu} \approx 0.6$ for a fully ionized gas, and $\tilde{\mu}\approx 1.3$ for a neutral gas.

The phase of the ISM in which the shock evolves is especially important, since the criterion of Eq.~\ref{eq:R_1} depends strongly on the ionization fraction, $x$.
Typically, SNR shocks are expected to be expanding in the warm ionized medium (WIM, $T\approx 8000$ K, $n_0 \approx 0.2-0.5$ cm$^{-3}$, volume filling factor $\sim 20-50$, $x=0.6-1$) or in the warm neutral medium (WNM, $T\approx 8000$ K, $n_0 \approx 0.2-0.5$ cm$^{-3}$, volume filling factor $\sim 10-20$, $x=0.01-0.05$)~\citep{mihalas1981}. In the WNM, the low ionization fraction makes the condition of Eq.~\ref{eq:R_1} soon after the shock enters the radiative phase $\sim 20-40$ kyr, whereas the ionization fraction found in the WIM, potentially close to 1, relaxes this condition. Moreover, at shocks (and especially  SNR shocks) several radiation fields can ionize the ISM upstream and downstream: this is the case for external radiation fields~\citep{sutherland2017}, as well as self-irradiated shocks, which emit in the UV and X-ray domain and have been shown, even for low-velocity shocks ($\sim$ a few $10$ km/s) to substantially affect the ionization fraction, as well as the thermodynamical structure of the shock~\citep{sarkar2021,godard2024,godard2024_2}. 
In addition to the shock structure modification, and ion-neutral damping, DSA itself is expected to be modified intrinsically at collisionless shocks propagating in partially ionized media~\citep{morlino2013}. A high level of ionization is required for DSA, and we thus assume that either the phase of the ISM in which the shock expands (e.g., in the WIM phase) or other additional sources, provide the required conditions for ionization.

\section{Supernova remnants}
\label{sec:SNR}

\subsection{Dynamics of supernova remnant shocks}
For a typical remnant from a Type Ia SN (thermonuclear), expanding in a homogeneous ISM, the time evolution of the shock radius, $r_{\rm sh}$, and velocity, $v_{\rm sh}$, in the FE and Sedov-Taylor (ST) phases is well described by self-similar solutions that have been derived in several works~\citep{chevalier1982,truelove1999,ptuskin2005}. For instance, in the adiabatic ST phase, 
\begin{align}
    r_{\rm sh} = 4.3 \left( \frac{E_{51}}{n}\right)^{1/5} t_{\rm kyr}^{2/5} \text{pc} \\
    v_{\rm sh} = 1.7 \times 10^{8} \left( \frac{E_{51}}{n}\right)^{1/5} t_{\rm kyr}^{-3/5} \text{cm/s}
.\end{align}
The adiabatic phase typically ends at $t_{\rm  ST}$ when the cooling time is on the order of the age of the SNR. 
Considering the prescription of~\citet{chevalier1994}, this leads to
\begin{equation}
t_{\rm  ST} \approx 20 \;   E_{\rm 51}^{0.35}{n}^{-0.57}  \text{kyr}
.\end{equation}
Considering the prescription of \citet{draine2011} for the cooling function leads to a $t_{\rm  ST}$ a factor of 2 longer.
When the shock enters the radiative phase, the shock velocity and radius scale as $v_{\rm sh}\propto t^{-5/7}$ and $r_{\rm sh} \propto t^{2/7}$~\citep{cioffi1988,bandiera2004,gintrand2020}. Other works, relying for example on 1D or 2D simulations, have also found a power-law index closer to $0.33$~\citep{blondin1998}.
The duration of the radiative phase can be estimated by requiring the shock velocity to remain a few times larger than the sound speed in the ISM, i.e.,  $v_{\rm sh} \gtrsim 4-5 c_{\rm s}$ with $c_{\rm s}=\left(\gamma k_{\rm B}T/m \right)^{1/2}\approx 10 T_4^{1/2}$ km/s with $T= T_4 10^{4}$K. This condition also ensures that the shock velocity remains super-Alfvènic, and leads to an estimate of the duration of the radiative phase:
\begin{equation}
t_{\rm  rad} \approx 3 \; 10^2 \;  n^{-0.37}   E_{\rm 51}^{0.35} T_{4}^{-0.7} \text{kyr}
.\end{equation}

The SNRs from a Type II (core--collapse) SN typically expand into a complex medium shaped by the evolution of their massive progenitor stars. During the main sequence, fast stellar winds carve out a low-density, hot bubble in pressure equilibrium with the ISM. In the red supergiant (RSG) phase, a slower, denser wind develops. After the SN explosion, the shock propagates through this dense wind, then the hot bubble, and finally into the ISM. The RSG wind density is 
$n_w = \frac{\dot{M}}{4\pi m u_w r^2}$, with $ \dot{M} \sim 10^{-5} \, M_\odot/\text{yr}$,   $u_w \sim 10^6 $ cm/s
 and $ m \sim 1.27  m_{\rm p}$ the mean mass per hydrogen nucleus. The bubble density is $n_{\rm b} = 0.01 \left(L_{36}^6 n_0^{19} t_{\text{Myr}}^{-22} \right)^{1/35} \, \text{cm}^{-3}$, following \citet{weaver1977}, with a typical main-sequence lifetime of  $\sim \text{Myr}$ \citep{longair1994}. The transition radius between the RSG wind and the hot bubble is typically located at a distance of $r_1 = \frac{\dot{M} u_w}{4\pi k n_b T_b}$, and the hot bubble radius is $r_b = 27 \left( \frac{L_{36}}{1.27 n_0} \right)^{1/5} t_{\text{Myr}}^{3/5} \, \text{pc}$.
Due to the structured medium, self-similar solutions are not easily found. However, under the thin-shell approximation -- i.e., assuming that the swept-up material accumulates in a thin shell -- an analytical treatment is still possible \citep[e.g.,][]{ostriker1988,bisnovatyikogan1995}. For a spherically symmetric case, \citet{ptuskin2005} derived expressions for the shock velocity, $ v_{\text{sh}}(R_{\text{sh}})$, and age, $t(R_{\text{sh}})$. For typical type II SNe, we adopted $ E_{\text{SN}} = 10^{51}$ erg, $M_{\text{ej}} = 5\, M_{\odot} $, and $ \dot{M} = 10^{-5}\, M_\odot/\text{yr}$.

Throughout the evolution of the SNR shock, the magnetic field strength at the shock front undergoes significant variations. In the early stages, observations of X-ray filaments provide clear evidence that the magnetic field is substantially amplified compared to typical ISM values~\citep{vink2012}. Several mechanisms have been proposed to account for this amplification at the shock. At young SNR shocks (in the FE and early ST phase), the main mechanism responsible for magnetic field amplification is expected to be due to the streaming of CRs' excited nonresonant instabilities in the plasma upstream as the stream from the shock~\citep{bell2004,schure2013,bell2013}, often referred to as the Bell mechanism, which typically lead to a magnetic field of $B\propto \rho^{1/2} u_{\rm sh}^{3/2}$ (see e.g.,  Eq. 12 in~\citet{cristofari2021}).  To account for our incomplete knowledge of the problem, and the fact that other mechanisms might come into play and complicate the picture, several groups have proposed various physically motivated prescriptions, where the magnetic field is parametrized to account for nonlinear effects as in~\citet{morlino2012,diesing2019}, leading to $B \propto \sqrt{\rho} u_{\rm sh}$, and referred to it as the “resonant modified” case (see discussion and  Eq. 5-11 in~\citet{cristofari2021}).
Here we work under the usual assumption the nonresonant (Bell) streaming instability sets the value of the amplified magnetic field, while the conditions for the excitation of these modes are met.

\begin{figure}
    \centering
	\includegraphics[width=2.9in]{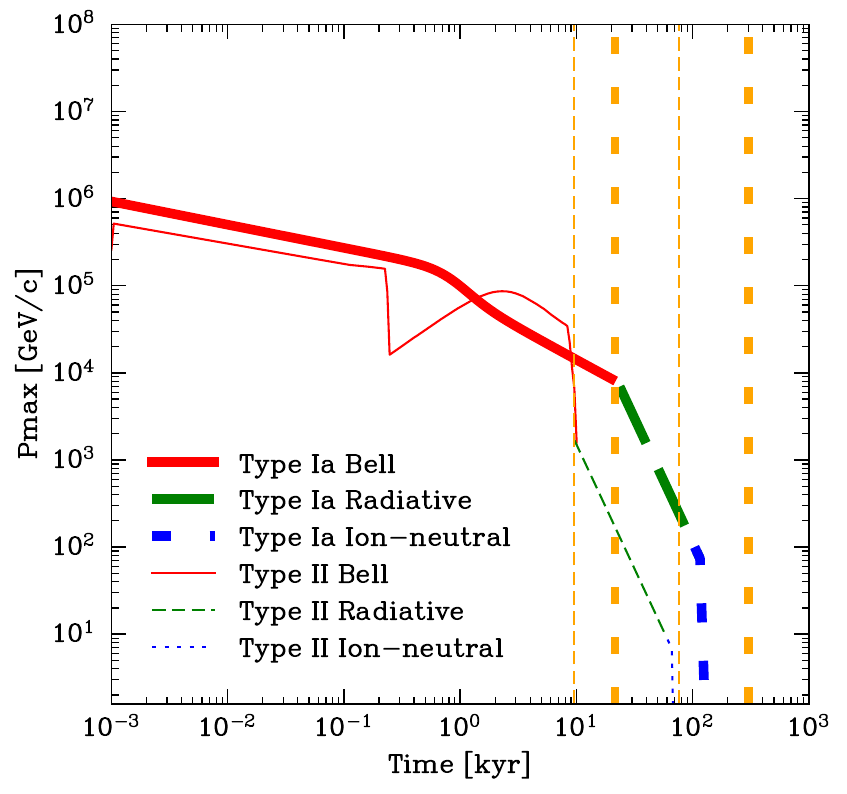}
    \caption{Maximum momentum of accelerated particles at an SNR from a Type Ia (thick lines) and Type II  (thin) progenitor. The dashed orange and black lines correspond to the beginning and end of the radiative phase. The solid red lines correspond to the $p_{\rm max}$ associated with a magnetic field amplification driven by the growth of nonresonant streaming instabilities (Bell)~\citep{schure2013}, the dashed green lines to the radiative phase (Section~\ref{sec:pmax}), and the dotted blue line to the ion-neutral damping (Section~\ref{sec:damping}). }
    \label{fig:pmax}
\end{figure}

 In addition to these prescriptions, we imposed in our estimate of $p_{\rm max}$  the conditions discussed in Sec.~\ref{sec:pmax} that limit the maximum energy when the shock becomes radiative.  In the radiative phase of typical SNRs, the conditions for excitations of instabilites that lead to magnetic field amplification  are not met, and the magnetic fied is thus assumed to remain unamplified upstream, and compressed downstream by a factor of $\sqrt{(1+2r^2)/3}$.
 Overall, the maximum momentum obtained is illustrated for a typical Type Ia SNR shock in Fig.~\ref{fig:pmax}, and is typically found between a few times $\sim 10^{3}$ GeV/c and $\sim 1$ GeV/c in the radiative phase. For the Type II, in the radiative phase, $p_{\rm max} \lesssim 10^3$ GeV/c. For both types, the limitation of $p_{\rm max}$ due to ion-neutral damping becomes important before the naïve estimate of the end of the radiative phase, typically $\sim 150$ kyr (Type Ia) and $\sim 60 kyr$ (Type II); also, these values are not well constrained (see discussion in Sec.~\ref{sec:pmax} and Sec.~\ref{sec:SNR}).

\subsection{The proton and electron content}
We computed the spectrum of protons accelerated throughout the life of the SNR, up to the end of the radiative phase, under the assumption that particle acceleration proceeds in the radiative phase as was discussed in the previous section. 
The calculation was carried out as in~\citet{cristofari2020,cristofari2021}, and we refer the reader to~\citet{cristofari2021} for the analytical expressions. Until the end of the radiative phase, the approach distinguishes between the particles accelerated and trapped inside the SNR suffering adiabatic losses (and radiative losses for electrons), and the particles at the highest energy continuously escaping in the ISM. In addition, the protons and electrons reaccelerated from Galactic CRs were computed. 
These contributions can be estimated by assuming that the protons and electrons in front of the SNR shocks are the ones of the local interstellar spectrum, parametrized by~\citep{bisschoff2019}, in agreement with the Voyager I~\citep{cummings2016} and PAMELA data~\citep{adriani2011}.

At the shock, the spectrum of reaccelerated particles is 
\begin{equation}
f_0^{\rm reac} = s \int_{p_0}^p \frac{\dd p'}{p'} \left( \frac{p'}{p} \right)^s f_{\infty}(p')
,\end{equation}

with $s= 3r/(r-1)$ with $r$ the compression factor felt by reaccelerated particles. 
 The associate cumulative spectrum of reaccelerated particles reads

 \begin{equation}
\label{eq:number}
N^{\rm reac}_{\rm loss}(p) = \int_{t_0}^{T_{\rm rad}}  \textrm{d}t \frac{4 \pi}{r} R_{\rm sh}^2(t) \;  v_{\rm sh}(t) \left(\frac{p'}{p}\right)^2\; f^{\rm reac}_0(p',t) \frac{\text{d}p'}{\text{d}p} 
,\end{equation}
 
 This is computed assuming that the change in momentum of a particle injected at a time, $t'$, with momentum, $p'$, due to losses can be written as 
 \begin{equation}
\frac{\text{d}p}{\text{d}t}= - \frac{p}{\cal L} \frac{\text{d} \cal L}{\text{d}t} - \frac{4}{3} \sigma_{\rm T}  \left( \frac{p}{m_{\rm e} c} \right)^2 \frac{B_2^2(t)}{8 \pi},
\label{eq:change}
\end{equation}
where $\sigma_{\rm T}$ is the Thomson cross section. ${\cal L}$ accounts for adiabatic energy losses, in terms of a change in volume between the two times, $t'$ and $t$: 
 \begin{equation}
{\cal L} (t,t')= \left( \frac{\rho_{\rm down}(t)}{\rho_{\rm down}(t')}\right)^{1/3},
\label{eq:conservation}
\end{equation}
with $\rho_{\rm down}$ the density downstream. If the expansion is adiabatic, $\rho_{\rm down}\propto P^{1/\gamma} \propto (\rho v_{\rm sh}^2(t))^{1/\gamma}$, and $\rho$ is the gas density upstream of the shock. For protons, synchrotron losses are negligible, while for electrons both adiabatic and radiative losses are important.

\section{Results}

\subsection{The cumulative proton and electron spectra}

\begin{figure*}
    \centering 
	\includegraphics[width=2.9in]{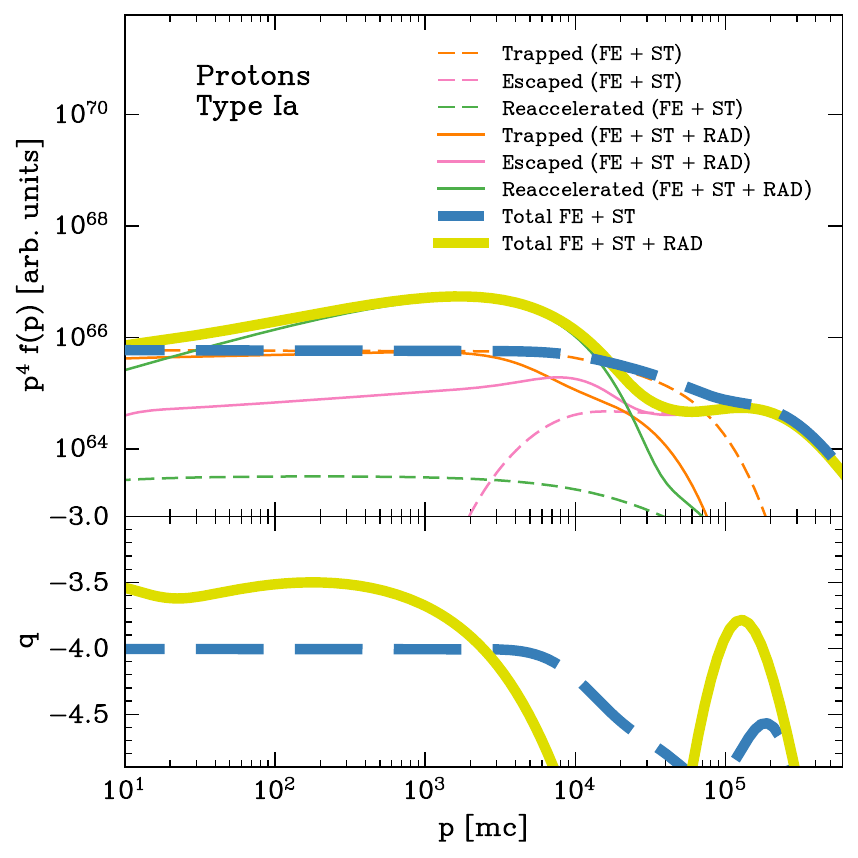}
		\includegraphics[width=2.9in]{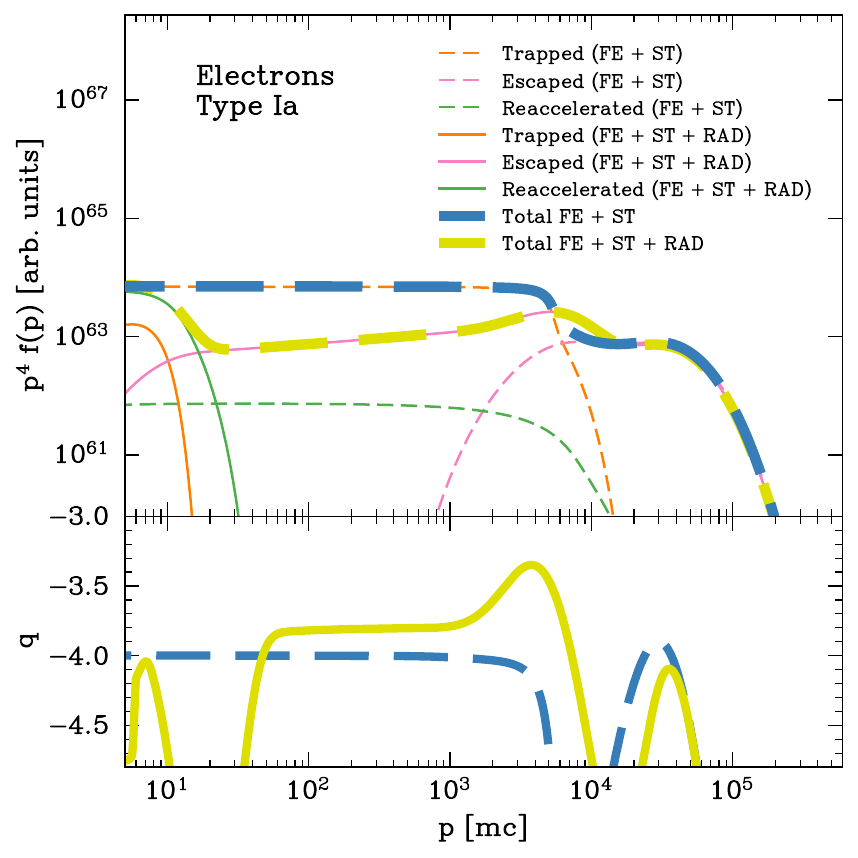}
    \caption{Protons (left panel) and electrons (right panel) accelerated throughout the life of an SNR from a Type Ia SN. The thin orange, pink, and green lines correspond to the particles trapped inside the SNR, escaping the SNR, and reaccelerated from the Galactic CRs,  respectively. The dashed lines correspond to the FE and ST phases, while the solid lines additionally take into account the radiative (RAD) phase.
     The thick dashed blue line corresponds to the total contribution of the FE and ST phases. The thick solid lime line corresponds to the total contribution of the FE, ST, and RAD phases. 
    The bottom panels show the slope of the spectrum sum of all components, for the ST and FE phase (dashed blue), and including the radiative phase (solid lime).}
    \label{fig:SNR_typeIa}
\end{figure*}

\begin{figure*}
    \centering
	\includegraphics[width=2.9in]{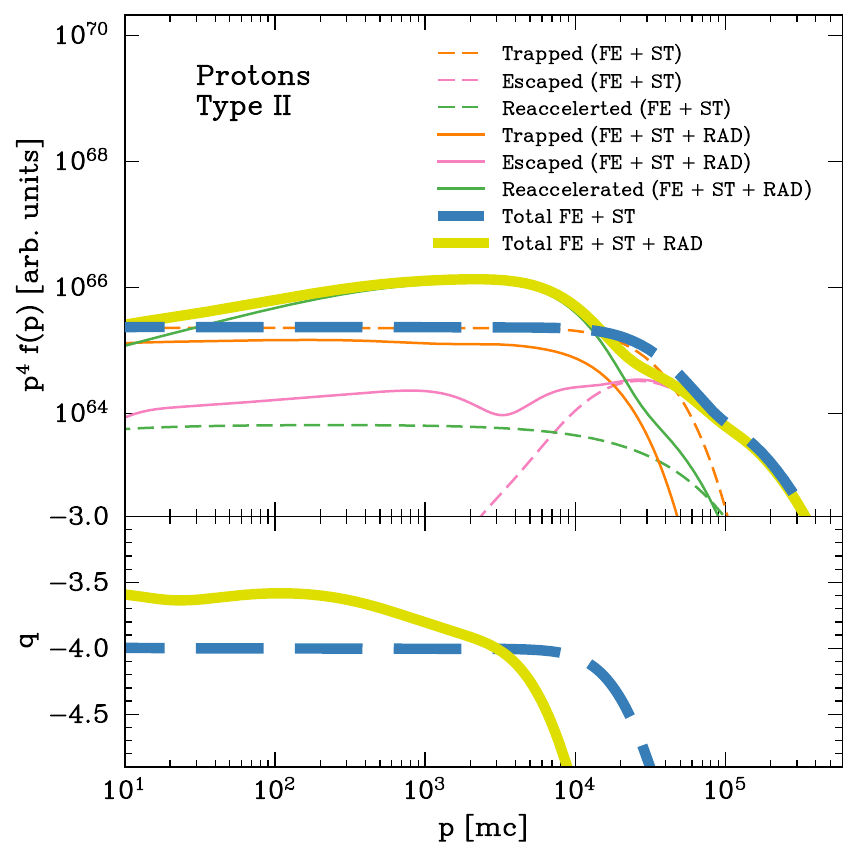}
		\includegraphics[width=2.9in]{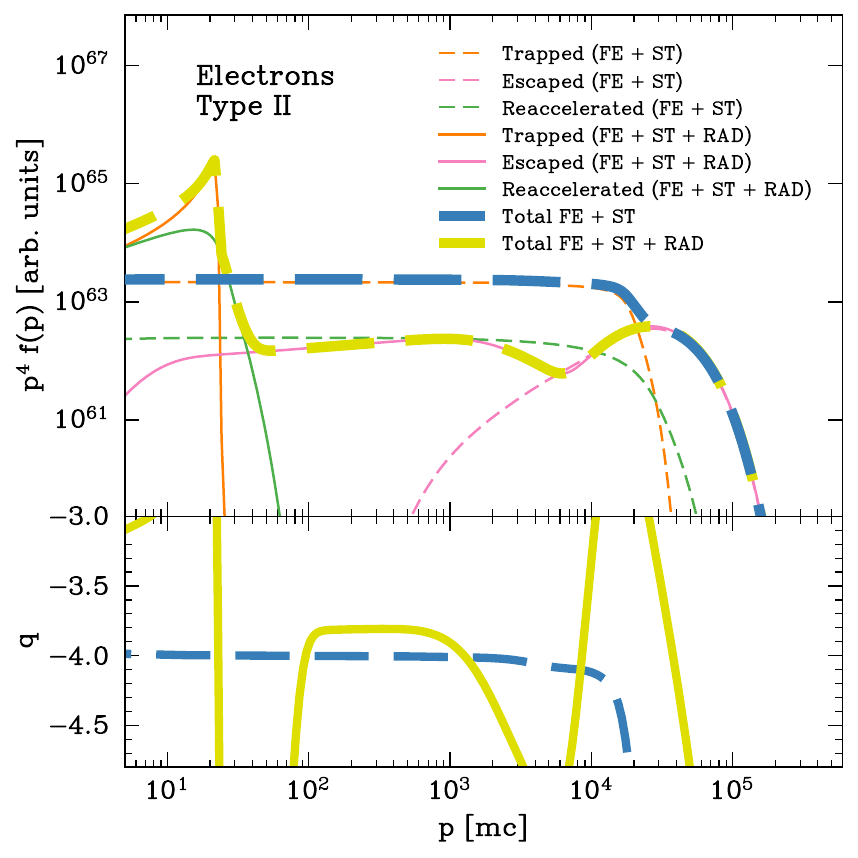}
    \caption{As in Fig.~\ref{fig:SNR_typeIa}, but for a SNR from a Type II SN, $E_{\rm SN}=10^{51}$erg, $\dot{M}=10^{-5}$ M$_{\odot}$/yr, and $M_{\rm ej}=5$ M$_{\odot}$.}
    \label{fig:SNR_typeII}
\end{figure*}

The total proton and electron spectra at SNRs from the Type Ia and Type II prototypes were computed as in~\citet{cristofari2021}, with the main additions being the extension to the radiative phase discussed in Sec.~\ref{sec:rad} and Sec.~\ref{sec:SNR} to account for the spectrum and maximum momentum of accelerated particles at the radiative shock, and for the reacceleration of preexisting CRs. As is shown in Fig.~\ref{fig:SNR_typeIa} and  Fig.~\ref{fig:SNR_typeII}, from 10 GeV to 10 TeV the inclusion of the radiative phase  produces a “bump” in the proton spectra, i.e., harder than $p^{-4}$ below a few tera-electronvolts and steeper above. This is mostly due to the component from reaccelerated protons, which scales with the volume of the SNRs (thus becoming more important with time), and the fact that at later times the spectra of accelerated particles become harder (closer to $p^{-3}$) as the shock becomes radiatiave. For electrons, the radiative losses drastically affect the particles trapped inside the SNRs (both accelerated and reaccelerated), so that the dominant component is due to the escaping particles. Several bumps appear, due to the evolution of the maximum energy of accelerated electrons throughout the ST and radiative phases. At low energy (around 10 GeV), the pile-up of trapped particles that cooled radiatively contributes to the total spectrum.

\subsection{The  electron--proton spectral index difference}
\begin{figure}[h]
\centering
	\includegraphics[width=2.9in]{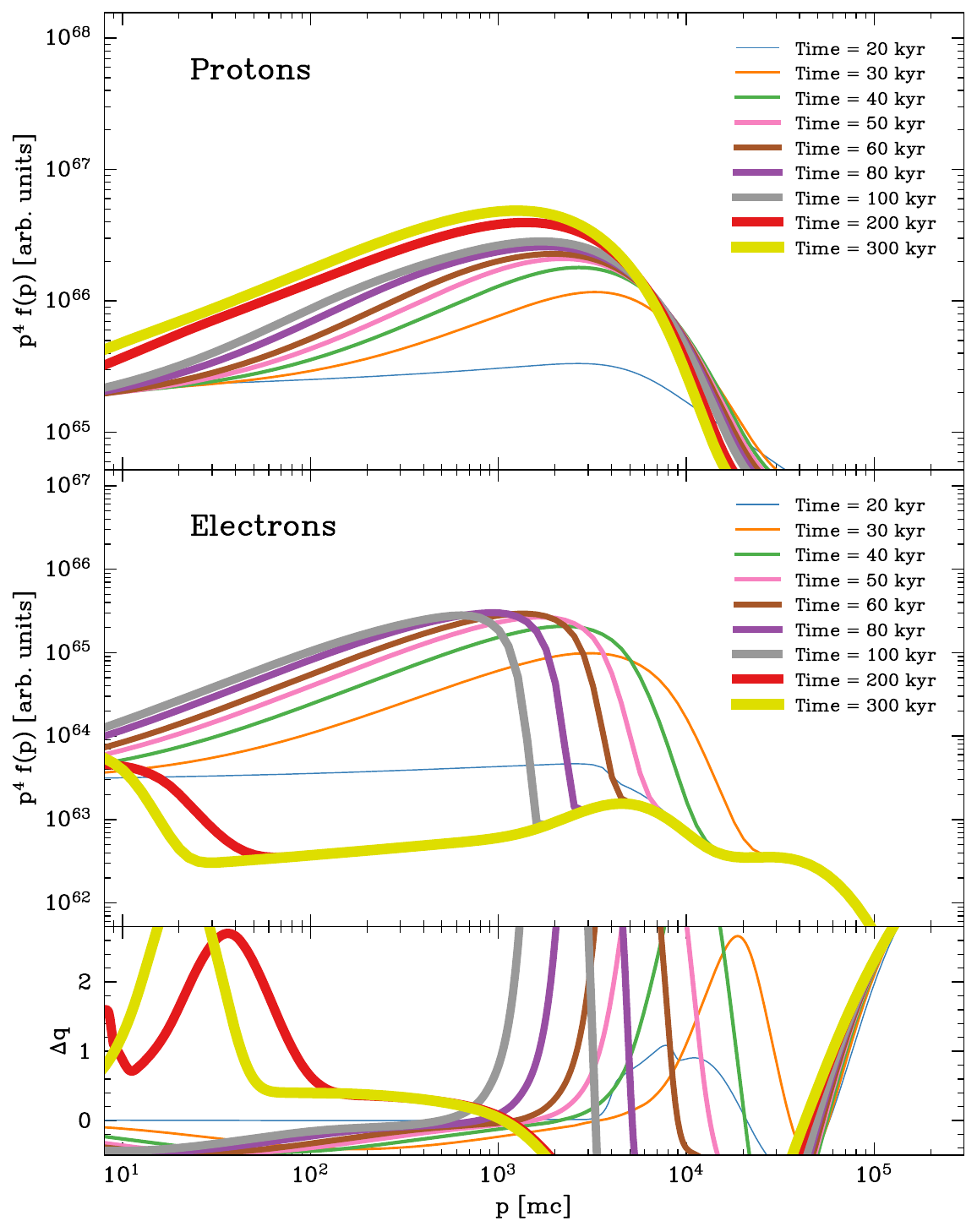}
    \caption{Proton (top), electron (middle), and proton--electron spectral index difference,  $\Delta q= q_{\rm electrons}- q_{\rm protons}$ (bottom), distributions. Colored lines (from thin to thick) show the effect of increasing the SNR lifetime from 20 kyr to 300 kyr, assuming a Type Ia SN progenitor.}
            \label{fig:time}
\end{figure}

\begin{figure}[h]
\centering
	\includegraphics[width=2.9in]{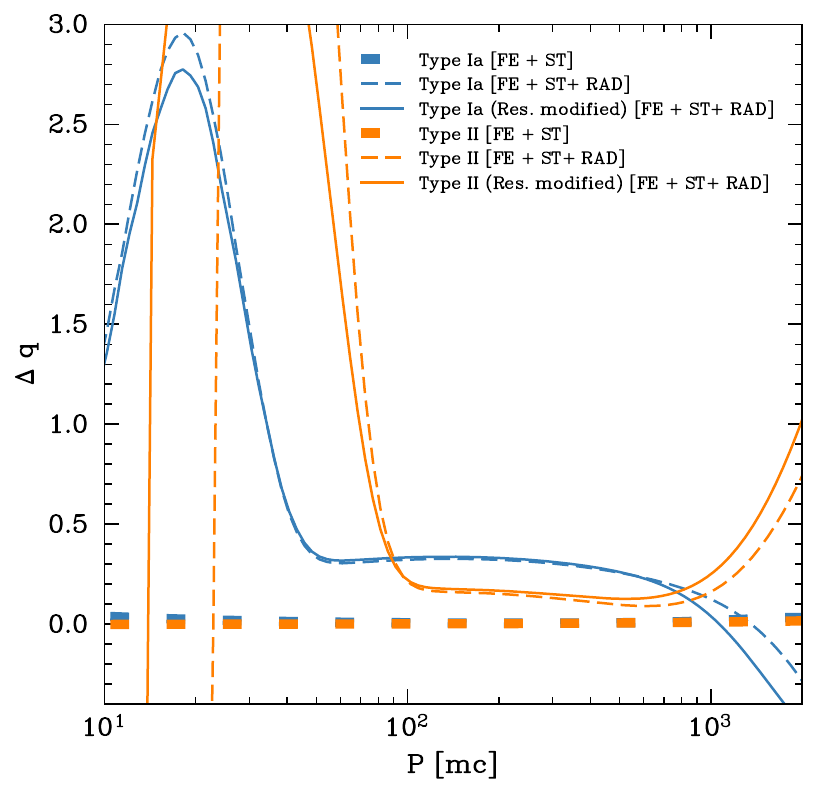}
    \caption{Electron–proton spectral index difference, $\Delta q= q_{\rm electrons}- q_{\rm protons}$, for Type Ia (blue lines) and Type II (orange lines). The thick lines correspond to the results considering the FE and ST phases, the thin solid lines additionally include the radiative phase considering Bell prescription for the magnetic field,  and the dashed lines considering the resonant modified prescription for the magnetic field. }
            \label{fig:electronproton}
\end{figure}

During the radiative phase of the SNR, particle trapping can significantly increase the radiative losses experienced by electrons. As is shown in Fig.~\ref{fig:time}, at times $\gtrsim 100$ kyr this tends to make the electron spectra steeper than in the earlier phases of the SNR evolution. Assuming that the accelerated protons and electrons trapped inside the SNR remain confined until the end of the radiative phase ($\sim 300$ kyr for a Type Ia, and $\sim 70$ kyr for a Type II), the electron spectra are found to be significantly steeper than the proton spectra for both type, as is illustrated in Fig.~\ref{fig:electronproton}, with a typical difference at $10^3 \gtrsim p \gtrsim 10^2$ mc of $\sim 0.1-0.2$ (Type II), or $\sim 0.3$ (Type Ia). 
Considering a different prescription for the magnetic field, for instance the resonant modified case discussed in Sec.~\ref{sec:SNR}, does not significantly affect our results. 
In our calculation, at $p \lesssim 10^2$ mc, the electron–proton spectral index difference becomes large due to the accumulation of electrons that have cooled via radiative losses. However, this part of the spectrum should be interpreted with caution, given the current uncertainties on particle confinement efficiency and the actual role of reacceleration at SNR shocks, especially late in the radiative phase.

\subsection{Discussion}
Several assumptions should be kept in mind, as they could affect the proton and electron spectra presented in the previous section, although they are not expected to alter the overall conclusions of this study.
 First, we worked under the assumption that the trapped particles, whether accelerated or reaccelerated, are trapped until the end of the radiative phase, suffering adiabatic and radiative losses. This is of course a strong assumption, as it is likely that in time the trapping and confinement of the accelerated particles could become inefficient, and thus a fraction of these particles could start leaking in the ISM. 
In addition, we worked under the assumption that as the SNR shocks expand, the magnetic field downstream is advected without suffering any damping~\citep{pohl2005,marcowith2010,ressler2014,tran2015,wilhelm2020}. Such damping has been proposed in several works, and could be especially important for the radiative losses suffers by electrons, which typically scale with $\propto B^2$. 

The duration of the radiative phase and/or the onset of the ion-neutral damping or other effects that could completely prevent particle acceleration through DSA are still not well constrained by observations and theoretical works, and the estimates presented in this work  have to be taken with caution. Fig~\ref{fig:time} illustrates the importance of the duration of the radiative phase in which DSA can take place. 
 For instance, in the case of efficient trapping, a radiative phase longer than $\gtrsim 100$ kyr is required to ensure that the electron-proton spectral difference is $\Delta q >0.1$. If the confinement of trapped particles is only partially efficient, the cumulative electron spectrum may steepen more rapidly, causing $\Delta q$ to reach values of $0.2-0.3$ at earlier times. In this work, we assumed that at strong shocks the instantaneous spectra of accelerated particles follow $\propto p^{-4}$ (i.e. $\propto E^{-2}$ for $E \gg 1$ GeV), neglecting nonlinear effects that may introduce deviation in the accelerated spectra (e.g., due to efficient particle acceleration~\citep{malkov2001,amato2005}, or due to the drift of scattering centers upstream or downstream~\citep{zirakahsvili2008,caprioli2020,cristofari2022}. These effects were included for a more complete description of the cumulative spectrum, but overall, since they affect electrons and protons at the shock in the same manner, they do not impact our result for $\Delta q$.

The idea that the radiative phase contributes to the total spectrum is compatible with the fact that all SNR shocks in the radiative phase have not been detected as high-energy particle accelerators (via radio or gamma-ray observations). Indeed, in the radiative phase, the shock velocity, $v_{\rm sh}$, is rather low (few 100 km/s), and thus the available ram pressure ($\rho v_{\rm sh}^2$) is substantially smaller than in the FE or ST phase; thus, at any given time, the instantaneous amount of particle accelerated is substantially smaller than in the  FE or ST phase, and the associated emission from nonthermal particles is below telescopes' sensitivities. 
In the approach adopted here, Type Ia SNR shocks can typically contribute up to $\sim 200$ kyr, leading to a typical shock radius of $r_{\rm sh}\sim 35$ pc. This suggests that very low-velocity SNRs expand in the ISM and have not been detected through multiwavelength observations. This is somewhat expected in the sense that in addition to the decrease in the ram pressure, $\propto v_{\rm sh}^{-10/7}$, and thus of the typical energy density of accelerated particles at the shock, the angular extension of such low-velocity evolved shocks would make their detection highly challenging. 

The Galactic SN rate is often claimed to be $\sim 3$/century (with values estimated from various methods found between $\approx 2.5/$century and $\approx 5.7/$century ~\citep{strom1994,tammann1994,smartt2009,adams2013}). As was discussed above, the lifetime of  SNRs can vary substantially from one SN to the other, but typically, if the end of the radiative phase happens at $\sim 100$ kyr, this means that currently there should be $\sim 3000$ SNR in the Galaxy, when radio surveys have so far revealed $\approx 310$ SNRs~\citep{green2025}, suggesting that 90 \% of the Galactic SNRs would remain undetected. Those would a priori be the oldest, radiative, and most extended SNRs.

Furthermore, the idea that a substantial fraction of the radiative phase can contribute to particle acceleration up to $\gtrsim 10^2-10^3$ GeV is also in favor of a reduced CR efficiency, closer to $\sim 0.01$ of the ram pressure than the $\sim 0.1$ often used to fit gamma-ray observations of SNRs~\citep{HESSSNR2018}, or to account for gamma-ray observations of the Galactic plane~\citep{acero2016}. Thus if the radiative phase indeed contributes to particle acceleration, such an increase in the duration of the accelerating phase implies a reduction of the efficiency of acceleration.

\subsection{Very young radiative supernova shocks}

Core-collapse SNe typically explode into the dense wind of a late-stage massive star, characterized by a mass-loss rate, $\dot{M}$, and wind velocity, $u_{\rm w}$, leading to a density profile of $\rho = \dot{M}/(4\pi u_{\rm w} r^2)$. The high density enhances radiative cooling, thereby reducing the maximum momentum attainable by accelerated particles. During the late stages of massive star evolution, the mass-loss rate can vary substantially, and a wide range of values is reported in the literature. Depending on the ejecta mass and the total explosion energy of the SN, the resulting shock velocity in the wind is well described by self-similar solutions~\citep{tang2016}.  

The case of very young SN shocks (from days to about a year after the explosion) is of particular interest because, unlike the typical radiative SNR shocks ($\gtrsim 20$ kyr), the shock speed and density are sufficiently high to excite Bell instabilities in the plasma, leading to significant amplification of the magnetic field~\citep{marcowith2018}. At the same time, the high density can make radiative cooling important. These conditions combine properties usually associated with both young, non-radiative SNRs and old, radiative SNRs, providing a unique environment in which to study the interplay between magnetic field amplification, particle acceleration, and radiative losses.

\begin{figure}
    \centering
	\includegraphics[width=2.9in]{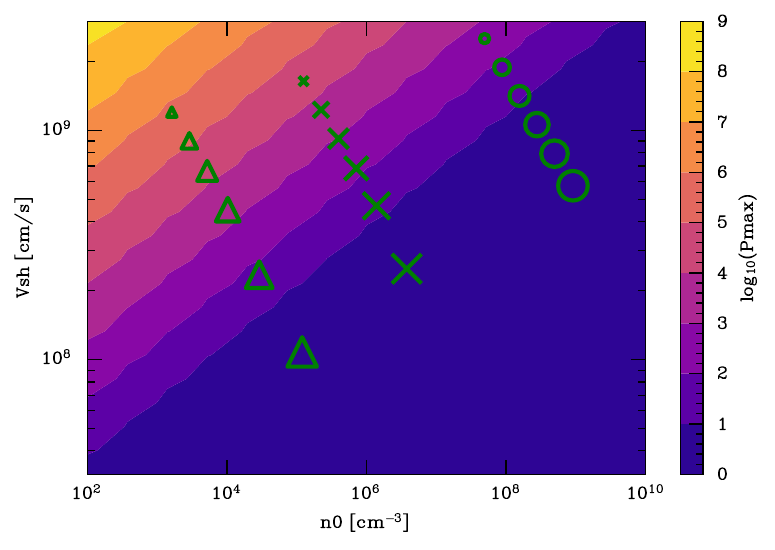}
    \caption{Maximum momentum of accelerated particles at a young SNR shock expanding in the dense wind of the progenitor star, as a function of density and shock velocity. The green circles, crosses, and triangles show the typical values reached 1 day, 1 month, and 1 year after the SN explosion, respectively. The size of the symbols increases with $\dot{M}$ from $10^{-5}$ to 1 M$_{\odot}$/yr.}
    \label{fig:Emax_contour_young_SN.pdf}
\end{figure}

The maximum energy of accelerated particles, estimated with the criteria described in Sec.~\ref{sec:pmax}, is illustrated in Fig.~\ref{fig:Emax_contour_young_SN.pdf}. On the first day, for mass-loss rates of $\dot{M} \sim 10^{-2}-1$ M$_{\odot}$/yr, cooling suppresses acceleration. At one month and one year, this limitation persists only for mass-loss rates on the order of $\dot{M} \sim 1$ M$_{\odot}$/yr. For lower mass-loss rates, $\dot{M} \lesssim 10^{-3}$ M$_{\odot}$/yr, particle energies in the tera-electronvolt range can typically be reached at these timescales, opening up interesting prospects for detection with next-generation observatories~\citep{CTAScience,LHAASO2024}.

\section{Conclusions}
Radiative cooling downstream of astrophysical shock waves significantly impacts the shock structure --affecting temperature, velocity, and density -- and, consequently, influences the potential for particle acceleration through DSA. In the context of SNRs, the contribution of the radiative phase to particle acceleration is often overlooked, as it is typically considered subdominant compared to the free-expansion and adiabatic phases. However, our results suggest that the radiative phase can have a substantial impact on the total particle spectra. For protons, the cumulative spectra are found to be harder than the canonical test-particle spectrum, $p^{-4}$, for $p \lesssim 10^3$ GeV/c, and steeper above this energy. The electron spectra are also affected, showing several bumps and deviations from $p^{-4}$, mainly due to the combined effects of the time evolution of the maximum electron momentum and the losses suffered by confined electrons.

The contribution of the radiative phase leads to an electron–proton spectral index difference of $\Delta q \sim 0.1$–0.2 (Type II) and $\Delta q \sim 0.3$ (Type Ia) for $10^3 \gtrsim p \gtrsim 10^2$ GeV/c, which can be of great interest in the search for the sources of Galactic CRs~\citep{blasi2013,gabici2019}. The spectral hardening/steepening and deviations from $p^{-4}$ due to the radiative phase should also be examined in the context of the precision measurements of protons by DAMPE or CALET~\citep{DAMPE_protons,CALET_protons} and other experiments, as they may play a role in shaping the local CR spectrum.

Several aspects of our study need to be examined more thoroughly to clarify the role of the radiative phase, including its duration, the damping of the downstream magnetic field, the maximum energy of accelerated particles, and the shock’s ability to confine particles within the SNR. Additionally, other aspects, such as instabilities in the cool shell that can substantially complicate  the picture and would need to be studied~\citep{chevalier1982b, binette1985}. 
Several theoretical works relying on HD simulations have discussed the complexity of the structure of radiative shocks~\citep{chevalier1982b}, as several instabilities can develop, such as the “nonlinear thin-shell instability,” which can deform the shock front, leading to a corrugated shock~\citep{vishniac1994,strickland1995}, and affect particle acceleration~\citep{steinberg2018}. 

The case of SNRs illustrates how low-velocity radiative shocks may play an important role in the Galactic ecosystem of accelerated charged particles. More generally, even if these shocks (such as old SNRs) remain so far often undetected, they can still contribute to injecting mechanical energy and electromagnetic turbulence in the ISM, thereby affecting not only the energization of particles but also the properties of the ISM that regulate the transport of CRs.In the coming years, the multiwavelength study of astrophysical accelerators across the electromagnetic spectrum, from next-generation radio observatories such as SKAO~\citep{LOFAR2013} to gamma-ray observatories such as CTAO~\citep{CTAScience}, will help clarify the role of radiative shocks in particle acceleration. 
Additionally, laboratory experiments with laser-generated radiative shocks have shown promising potential for studying particle acceleration in controlled settings~\citep{keilty2000,reighard2006,yao2021,sakawa2024}. 

\begin{acknowledgements}
PC acknowledges support from the "GALAPAGOS" PSL Starting Grant. PC thanks the anonymous referee for constructive comments, and is grateful to Pasquale Blasi, Benjamin Godard, Antoine Gusdorf, Guillaume Pineau des Forêts, and Guillaume Vigoureux for insightful discussions. 
\end{acknowledgements}

   \bibliographystyle{aa} % style aa.bst
   \bibliography{radiative} % your references Yourfile.bib

@ARTICLE{weaver1977,
       author = {{Weaver}, R. and {McCray}, R. and {Castor}, J. and {Shapiro}, P. and
         {Moore}, R.},
        title = "{Interstellar bubbles. II. Structure and evolution.}",
      journal = {\apj},
         year = "1977",
        month = "Dec",
       volume = {218},
        pages = {377-395},
          doi = {10.1086/155692},
       adsurl = {https://ui.adsabs.harvard.edu/abs/1977ApJ...218..377W}
}

@ARTICLE{bandiera2004,
       author = {{Bandiera}, R. and {Petruk}, O.},
        title = "{Analytic solutions for the evolution of radiative supernova remnants}",
      journal = {\aap},
         year = 2004,
        month = may,
       volume = {419},
        pages = {419-423},
          doi = {10.1051/0004-6361:20035950},
archivePrefix = {arXiv},
       eprint = {astro-ph/0402598},
 primaryClass = {astro-ph},
       adsurl = {https://ui.adsabs.harvard.edu/abs/2004A&A...419..419B}
}

@ARTICLE{CALET_protons,
       author = {{Adriani}, O. and {Akaike}, Y. and {Asano}, K. and {Asaoka}, Y. and {Berti}, E. and {Bigongiari}, G. and {Binns}, W.~R. and {Bongi}, M. and {Brogi}, P. and {Bruno}, A. and {Buckley}, J.~H. and {Cannady}, N. and {Castellini}, G. and {Checchia}, C. and {Cherry}, M.~L. and {Collazuol}, G. and {Ebisawa}, K. and {Ficklin}, A.~W. and {Fuke}, H. and {Gonzi}, S. and {Guzik}, T.~G. and {Hams}, T. and {Hibino}, K. and {Ichimura}, M. and {Ioka}, K. and {Ishizaki}, W. and {Israel}, M.~H. and {Kasahara}, K. and {Kataoka}, J. and {Kataoka}, R. and {Katayose}, Y. and {Kato}, C. and {Kawanaka}, N. and {Kawakubo}, Y. and {Kobayashi}, K. and {Kohri}, K. and {Krawczynski}, H.~S. and {Krizmanic}, J.~F. and {Maestro}, P. and {Marrocchesi}, P.~S. and {Messineo}, A.~M. and {Mitchell}, J.~W. and {Miyake}, S. and {Moiseev}, A.~A. and {Mori}, M. and {Mori}, N. and {Motz}, H.~M. and {Munakata}, K. and {Nakahira}, S. and {Nishimura}, J. and {de Nolfo}, G.~A. and {Okuno}, S. and {Ormes}, J.~F. and {Ozawa}, S. and {Pacini}, L. and {Papini}, P. and {Rauch}, B.~F. and {Ricciarini}, S.~B. and {Sakai}, K. and {Sakamoto}, T. and {Sasaki}, M. and {Shimizu}, Y. and {Shiomi}, A. and {Spillantini}, P. and {Stolzi}, F. and {Sugita}, S. and {Sulaj}, A. and {Takita}, M. and {Tamura}, T. and {Terasawa}, T. and {Torii}, S. and {Tsunesada}, Y. and {Uchihori}, Y. and {Vannuccini}, E. and {Wefel}, J.~P. and {Yamaoka}, K. and {Yanagita}, S. and {Yoshida}, A. and {Yoshida}, K. and {Zober}, W.~V. and {Calet Collaboration}},
        title = "{Observation of Spectral Structures in the Flux of Cosmic-Ray Protons from 50 GeV to 60 TeV with the Calorimetric Electron Telescope on the International Space Station}",
      journal = {\prl},
         year = 2022,
        month = sep,
       volume = {129},
       number = {10},
          eid = {101102},
        pages = {101102},
          doi = {10.1103/PhysRevLett.129.101102},
archivePrefix = {arXiv},
       eprint = {2209.01302},
 primaryClass = {astro-ph.HE},
       adsurl = {https://ui.adsabs.harvard.edu/abs/2022PhRvL.129j1102A}
}

@ARTICLE{DAMPE_protons,
       author = {{An}, Q. and {Asfandiyarov}, R. and {Azzarello}, P. and {Bernardini}, P. and {Bi}, X.~J. and {Cai}, M.~S. and {Chang}, J. and {Chen}, D.~Y. and {Chen}, H.~F. and {Chen}, J.~L. and {Chen}, W. and {Cui}, M.~Y. and {Cui}, T.~S. and {Dai}, H.~T. and {D'Amone}, A. and {De Benedittis}, A. and {De Mitri}, I. and {Di Santo}, M. and {Ding}, M. and {Dong}, T.~K. and {Dong}, Y.~F. and {Dong}, Z.~X. and {Donvito}, G. and {Droz}, D. and {Duan}, J.~L. and {Duan}, K.~K. and {D'Urso}, D. and {Fan}, R.~R. and {Fan}, Y.~Z. and {Fang}, F. and {Feng}, C.~Q. and {Feng}, L. and {Fusco}, P. and {Gallo}, V. and {Gan}, F.~J. and {Gao}, M. and {Gargano}, F. and {Gong}, K. and {Gong}, Y.~Z. and {Guo}, D.~Y. and {Guo}, J.~H. and {Guo}, X.~L. and {Han}, S.~X. and {Hu}, Y.~M. and {Huang}, G.~S. and {Huang}, X.~Y. and {Huang}, Y.~Y. and {Ionica}, M. and {Jiang}, W. and {Jin}, X. and {Kong}, J. and {Lei}, S.~J. and {Li}, S. and {Li}, W.~L. and {Li}, X. and {Li}, X.~Q. and {Li}, Y. and {Liang}, Y.~F. and {Liang}, Y.~M. and {Liao}, N.~H. and {Liu}, C.~M. and {Liu}, H. and {Liu}, J. and {Liu}, S.~B. and {Liu}, W.~Q. and {Liu}, Y. and {Loparco}, F. and {Luo}, C.~N. and {Ma}, M. and {Ma}, P.~X. and {Ma}, S.~Y. and {Ma}, T. and {Ma}, X.~Y. and {Marsella}, G. and {Mazziotta}, M.~N. and {Mo}, D. and {Niu}, X.~Y. and {Pan}, X. and {Peng}, W.~X. and {Peng}, X.~Y. and {Qiao}, R. and {Rao}, J.~N. and {Salinas}, M.~M. and {Shang}, G.~Z. and {Shen}, W.~H. and {Shen}, Z.~Q. and {Shen}, Z.~T. and {Song}, J.~X. and {Su}, H. and {Su}, M. and {Sun}, Z.~Y. and {Surdo}, A. and {Teng}, X.~J. and {Tykhonov}, A. and {Vitillo}, S. and {Wang}, C. and {Wang}, H. and {Wang}, H.~Y. and {Wang}, J.~Z. and {Wang}, L.~G. and {Wang}, Q. and {Wang}, S. and {Wang}, X.~H. and {Wang}, X.~L. and {Wang}, Y.~F. and {Wang}, Y.~P. and {Wang}, Y.~Z. and {Wang}, Z.~M. and {Wei}, D.~M. and {Wei}, J.~J. and {Wei}, Y.~F. and {Wen}, S.~C. and {Wu}, D. and {Wu}, J. and {Wu}, L.~B. and {Wu}, S.~S. and {Wu}, X. and {Xi}, K. and {Xia}, Z.~Q. and {Xu}, H.~T. and {Xu}, Z.~H. and {Xu}, Z.~L. and {Xu}, Z.~Z. and {Xue}, G.~F. and {Yang}, H.~B. and {Yang}, P. and {Yang}, Y.~Q. and {Yang}, Z.~L. and {Yao}, H.~J. and {Yu}, Y.~H. and {Yuan}, Q. and {Yue}, C. and {Zang}, J.~J. and {Zhang}, F. and {Zhang}, J.~Y. and {Zhang}, J.~Z. and {Zhang}, P.~F. and {Zhang}, S.~X. and {Zhang}, W.~Z. and {Zhang}, Y. and {Zhang}, Y.~J. and {Zhang}, Y.~L. and {Zhang}, Y.~P. and {Zhang}, Y.~Q. and {Zhang}, Z. and {Zhang}, Z.~Y. and {Zhao}, H. and {Zhao}, H.~Y. and {Zhao}, X.~F. and {Zhou}, C.~Y. and {Zhou}, Y. and {Zhu}, X. and {Zhu}, Y. and {Zimmer}, S.},
        title = "{Measurement of the cosmic ray proton spectrum from 40 GeV to 100 TeV with the DAMPE satellite}",
      journal = {Science Advances},
         year = 2019,
        month = sep,
       volume = {5},
       number = {9},
          eid = {eaax3793},
        pages = {eaax3793},
          doi = {10.1126/sciadv.aax3793},
archivePrefix = {arXiv},
       eprint = {1909.12860},
 primaryClass = {astro-ph.HE},
       adsurl = {https://ui.adsabs.harvard.edu/abs/2019SciA....5.3793A}
}

@ARTICLE{marcowith2010,
       author = {{Marcowith}, A. and {Casse}, F.},
        title = "{Postshock turbulence and diffusive shock acceleration in young supernova remnants}",
      journal = {\aap},
         year = 2010,
        month = jun,
       volume = {515},
          eid = {A90},
        pages = {A90},
          doi = {10.1051/0004-6361/200913022},
archivePrefix = {arXiv},
       eprint = {1001.2111},
 primaryClass = {astro-ph.HE},
       adsurl = {https://ui.adsabs.harvard.edu/abs/2010A&A...515A..90M}
}

@ARTICLE{tran2015,
       author = {{Tran}, Aaron and {Williams}, Brian J. and {Petre}, Robert and {Ressler}, Sean M. and {Reynolds}, Stephen P.},
        title = "{Energy Dependence of Synchrotron X-Ray Rims in Tycho's Supernova Remnant}",
      journal = {\apj},
         year = 2015,
        month = oct,
       volume = {812},
       number = {2},
          eid = {101},
        pages = {101},
          doi = {10.1088/0004-637X/812/2/101},
archivePrefix = {arXiv},
       eprint = {1509.00877},
 primaryClass = {astro-ph.HE},
       adsurl = {https://ui.adsabs.harvard.edu/abs/2015ApJ...812..101T}
}

@ARTICLE{ressler2014,
       author = {{Ressler}, Sean M. and {Katsuda}, Satoru and {Reynolds}, Stephen P. and {Long}, Knox S. and {Petre}, Robert and {Williams}, Brian J. and {Winkler}, P. Frank},
        title = "{Magnetic Field Amplification in the Thin X-Ray Rims of SN 1006}",
      journal = {\apj},
         year = 2014,
        month = aug,
       volume = {790},
       number = {2},
          eid = {85},
        pages = {85},
          doi = {10.1088/0004-637X/790/2/85},
archivePrefix = {arXiv},
       eprint = {1406.3630},
 primaryClass = {astro-ph.HE},
       adsurl = {https://ui.adsabs.harvard.edu/abs/2014ApJ...790...85R}
}

@ARTICLE{pohl2005,
       author = {{Pohl}, M. and {Yan}, H. and {Lazarian}, A.},
        title = "{Magnetically Limited X-Ray Filaments in Young Supernova Remnants}",
      journal = {\apjl},
         year = 2005,
        month = jun,
       volume = {626},
       number = {2},
        pages = {L101-L104},
          doi = {10.1086/431902},
       adsurl = {https://ui.adsabs.harvard.edu/abs/2005ApJ...626L.101P}
}

@ARTICLE{wilhelm2020,
       author = {{Wilhelm}, A. and {Telezhinsky}, I. and {Dwarkadas}, V.~V. and {Pohl}, M.},
        title = "{Stochastic re-acceleration and magnetic-field damping in Tycho's supernova remnant}",
      journal = {\aap},
         year = 2020,
        month = jul,
       volume = {639},
          eid = {A124},
        pages = {A124},
          doi = {10.1051/0004-6361/201936079},
archivePrefix = {arXiv},
       eprint = {2006.04832},
 primaryClass = {astro-ph.HE},
       adsurl = {https://ui.adsabs.harvard.edu/abs/2020A&A...639A.124W}
}

@BOOK{longair1994,
       author = {{Longair}, M.~S.},
        title = "{High energy astrophysics. Volume 2. Stars, the Galaxy and the interstellar medium.}",
         year = 1994,
       volume = {2},
       adsurl = {https://ui.adsabs.harvard.edu/abs/1994hea2.book.....L}
}

@ARTICLE{ptuskin2005,
       author = {{Ptuskin}, V.~S. and {Zirakashvili}, V.~N.},
        title = "{On the spectrum of high-energy cosmic rays produced by supernova remnants in the presence of strong cosmic-ray streaming instability and wave dissipation}",
      journal = {\aap},
         year = "2005",
        month = "Jan",
       volume = {429},
        pages = {755-765},
          doi = {10.1051/0004-6361:20041517},
archivePrefix = {arXiv},
       eprint = {astro-ph/0408025},
 primaryClass = {astro-ph},
       adsurl = {https://ui.adsabs.harvard.edu/abs/2005A&A...429..755P}
}

@ARTICLE{ostriker1988,
       author = {{Ostriker}, Jeremiah P. and {McKee}, Christopher F.},
        title = "{Astrophysical blastwaves}",
      journal = {Reviews of Modern Physics},
         year = "1988",
        month = "Jan",
       volume = {60},
       number = {1},
        pages = {1-68},
          doi = {10.1103/RevModPhys.60.1},
       adsurl = {https://ui.adsabs.harvard.edu/abs/1988RvMP...60....1O}
}

@ARTICLE{cummings2016,
       author = {{Cummings}, A.~C. and {Stone}, E.~C. and {Heikkila}, B.~C. and {Lal}, N. and {Webber}, W.~R. and {J{\'o}hannesson}, G. and {Moskalenko}, I.~V. and {Orlando}, E. and {Porter}, T.~A.},
        title = "{Galactic Cosmic Rays in the Local Interstellar Medium: Voyager 1 Observations and Model Results}",
      journal = {\apj},
         year = 2016,
        month = nov,
       volume = {831},
       number = {1},
          eid = {18},
        pages = {18},
          doi = {10.3847/0004-637X/831/1/18},
       adsurl = {https://ui.adsabs.harvard.edu/abs/2016ApJ...831...18C}
}

@ARTICLE{morlino2012,
       author = {{Morlino}, G. and {Caprioli}, D.},
        title = "{Strong evidence for hadron acceleration in Tycho's supernova remnant}",
      journal = {\aap},
         year = 2012,
        month = feb,
       volume = {538},
          eid = {A81},
        pages = {A81},
          doi = {10.1051/0004-6361/201117855},
archivePrefix = {arXiv},
       eprint = {1105.6342},
 primaryClass = {astro-ph.HE},
       adsurl = {https://ui.adsabs.harvard.edu/abs/2012A&A...538A..81M}
}

@ARTICLE{diesing2019,
       author = {{Diesing}, Rebecca and {Caprioli}, Damiano},
        title = "{Spectrum of Electrons Accelerated in Supernova Remnants}",
      journal = {\prl},
         year = 2019,
        month = aug,
       volume = {123},
       number = {7},
          eid = {071101},
        pages = {071101},
          doi = {10.1103/PhysRevLett.123.071101},
archivePrefix = {arXiv},
       eprint = {1905.07414},
 primaryClass = {astro-ph.HE},
       adsurl = {https://ui.adsabs.harvard.edu/abs/2019PhRvL.123g1101D}
}

@ARTICLE{petruk2024,
       author = {{Petruk}, O. and {Kuzyo}, T.},
        title = "{Individual particle approach to diffusive shock acceleration. Effect of the non-uniform flow velocity downstream of the shock}",
      journal = {\aap},
         year = 2024,
        month = aug,
       volume = {688},
          eid = {A108},
        pages = {A108},
          doi = {10.1051/0004-6361/202347803},
archivePrefix = {arXiv},
       eprint = {2404.17397},
 primaryClass = {astro-ph.HE},
       adsurl = {https://ui.adsabs.harvard.edu/abs/2024A&A...688A.108P}
}

@ARTICLE{malkov2001,
       author = {{Malkov}, M.~A. and {Drury}, L. O'C.},
        title = "{Nonlinear theory of diffusive acceleration of particles by shock waves}",
      journal = {Reports on Progress in Physics},
         year = 2001,
        month = apr,
       volume = {64},
       number = {4},
        pages = {429-481},
          doi = {10.1088/0034-4885/64/4/201},
       adsurl = {https://ui.adsabs.harvard.edu/abs/2001RPPh...64..429M}
}

@ARTICLE{adriani2011,
       author = {{Adriani}, O. and {Barbarino}, G.~C. and {Bazilevskaya}, G.~A. and {Bellotti}, R. and {Boezio}, M. and {Bogomolov}, E.~A. and {Bonechi}, L. and {Bongi}, M. and {Bonvicini}, V. and {Borisov}, S. and {Bottai}, S. and {Bruno}, A. and {Cafagna}, F. and {Campana}, D. and {Carbone}, R. and {Carlson}, P. and {Casolino}, M. and {Castellini}, G. and {Consiglio}, L. and {De Pascale}, M.~P. and {De Santis}, C. and {De Simone}, N. and {Di Felice}, V. and {Galper}, A.~M. and {Gillard}, W. and {Grishantseva}, L. and {Jerse}, G. and {Karelin}, A.~V. and {Koldashov}, S.~V. and {Krutkov}, S.~Y. and {Kvashnin}, A.~N. and {Leonov}, A. and {Malakhov}, V. and {Malvezzi}, V. and {Marcelli}, L. and {Mayorov}, A.~G. and {Menn}, W. and {Mikhailov}, V.~V. and {Mocchiutti}, E. and {Monaco}, A. and {Mori}, N. and {Nikonov}, N. and {Osteria}, G. and {Palma}, F. and {Papini}, P. and {Pearce}, M. and {Picozza}, P. and {Pizzolotto}, C. and {Ricci}, M. and {Ricciarini}, S.~B. and {Rossetto}, L. and {Sarkar}, R. and {Simon}, M. and {Sparvoli}, R. and {Spillantini}, P. and {Stozhkov}, Y.~I. and {Vacchi}, A. and {Vannuccini}, E. and {Vasilyev}, G. and {Voronov}, S.~A. and {Yurkin}, Y.~T. and {Wu}, J. and {Zampa}, G. and {Zampa}, N. and {Zverev}, V.~G.},
        title = "{PAMELA Measurements of Cosmic-Ray Proton and Helium Spectra}",
      journal = {Science},
         year = 2011,
        month = apr,
       volume = {332},
       number = {6025},
        pages = {69},
          doi = {10.1126/science.1199172},
archivePrefix = {arXiv},
       eprint = {1103.4055},
 primaryClass = {astro-ph.HE},
       adsurl = {https://ui.adsabs.harvard.edu/abs/2011Sci...332...69A}
}

@ARTICLE{bisschoff2019,
       author = {{Bisschoff}, D. and {Potgieter}, M.~S. and {Aslam}, O.~P.~M.},
        title = "{New Very Local Interstellar Spectra for Electrons, Positrons, Protons, and Light Cosmic Ray Nuclei}",
      journal = {\apj},
         year = 2019,
        month = jun,
       volume = {878},
       number = {1},
          eid = {59},
        pages = {59},
          doi = {10.3847/1538-4357/ab1e4a},
archivePrefix = {arXiv},
       eprint = {1902.10438},
 primaryClass = {astro-ph.HE},
       adsurl = {https://ui.adsabs.harvard.edu/abs/2019ApJ...878...59B}
}

@ARTICLE{bisnovatyikogan1995,
       author = {{Bisnovatyi-Kogan}, G.~S. and {Silich}, S.~A.},
        title = "{Shock-wave propagation in the nonuniform interstellar medium}",
      journal = {Reviews of Modern Physics},
         year = 1995,
        month = jul,
       volume = {67},
       number = {3},
        pages = {661-712},
          doi = {10.1103/RevModPhys.67.661},
       adsurl = {https://ui.adsabs.harvard.edu/abs/1995RvMP...67..661B}
}

@ARTICLE{bell2004,
       author = {{Bell}, A.~R.},
        title = "{Turbulent amplification of magnetic field and diffusive shock acceleration of cosmic rays}",
      journal = {\mnras},
         year = "2004",
        month = "Sep",
       volume = {353},
       number = {2},
        pages = {550-558},
          doi = {10.1111/j.1365-2966.2004.08097.x},
       adsurl = {https://ui.adsabs.harvard.edu/abs/2004MNRAS.353..550B}
}

@ARTICLE{guo2014a,
       author = {{Guo}, Xinyi and {Sironi}, Lorenzo and {Narayan}, Ramesh},
        title = "{Non-thermal Electron Acceleration in Low Mach Number Collisionless Shocks. II. Firehose-mediated Fermi Acceleration and its Dependence on Pre-shock Conditions}",
      journal = {\apj},
         year = 2014,
        month = dec,
       volume = {797},
       number = {1},
          eid = {47},
        pages = {47},
          doi = {10.1088/0004-637X/797/1/47},
archivePrefix = {arXiv},
       eprint = {1409.7393},
 primaryClass = {astro-ph.HE},
       adsurl = {https://ui.adsabs.harvard.edu/abs/2014ApJ...797...47G}
}

@ARTICLE{caprioli2014,
       author = {{Caprioli}, D. and {Spitkovsky}, A.},
        title = "{Simulations of Ion Acceleration at Non-relativistic Shocks. I. Acceleration Efficiency}",
      journal = {\apj},
         year = 2014,
        month = mar,
       volume = {783},
       number = {2},
          eid = {91},
        pages = {91},
          doi = {10.1088/0004-637X/783/2/91},
archivePrefix = {arXiv},
       eprint = {1310.2943},
 primaryClass = {astro-ph.HE},
       adsurl = {https://ui.adsabs.harvard.edu/abs/2014ApJ...783...91C}
}

@ARTICLE{park2015,
       author = {{Park}, Jaehong and {Caprioli}, Damiano and {Spitkovsky}, Anatoly},
        title = "{Simultaneous Acceleration of Protons and Electrons at Nonrelativistic Quasiparallel Collisionless Shocks}",
      journal = {\prl},
         year = 2015,
        month = feb,
       volume = {114},
       number = {8},
          eid = {085003},
        pages = {085003},
          doi = {10.1103/PhysRevLett.114.085003},
archivePrefix = {arXiv},
       eprint = {1412.0672},
 primaryClass = {astro-ph.HE},
       adsurl = {https://ui.adsabs.harvard.edu/abs/2015PhRvL.114h5003P}
}

@ARTICLE{cristofari2022,
       author = {{Cristofari}, Pierre and {Blasi}, Pasquale and {Caprioli}, Damiano},
        title = "{Microphysics of Diffusive Shock Acceleration: Impact on the Spectrum of Accelerated Particles}",
      journal = {\apj},
         year = 2022,
        month = may,
       volume = {930},
       number = {1},
          eid = {28},
        pages = {28},
          doi = {10.3847/1538-4357/ac6182},
archivePrefix = {arXiv},
       eprint = {2203.15624},
 primaryClass = {astro-ph.HE},
       adsurl = {https://ui.adsabs.harvard.edu/abs/2022ApJ...930...28C}
}

@ARTICLE{caprioli2020,
       author = {{Caprioli}, Damiano and {Haggerty}, Colby C. and {Blasi}, Pasquale},
        title = "{Kinetic Simulations of Cosmic-Ray-modified Shocks. II. Particle Spectra}",
      journal = {\apj},
         year = 2020,
        month = dec,
       volume = {905},
       number = {1},
          eid = {2},
        pages = {2},
          doi = {10.3847/1538-4357/abbe05},
archivePrefix = {arXiv},
       eprint = {2009.00007},
 primaryClass = {astro-ph.HE},
       adsurl = {https://ui.adsabs.harvard.edu/abs/2020ApJ...905....2C}
}

@INPROCEEDINGS{zirakahsvili2008,
       author = {{Zirakashvili}, Vladimir N. and {Ptuskin}, Vladimir S.},
        title = "{The influence of the Alfv{\'e}nic drift on the shape of cosmic ray spectra in SNRs}",
    booktitle = {American Institute of Physics Conference Series},
         year = 2008,
       editor = {{Aharonian}, Felix A. and {Hofmann}, Werner and {Rieger}, Frank},
       series = {American Institute of Physics Conference Series},
       volume = {1085},
        month = dec,
    publisher = {AIP},
        pages = {336-339},
          doi = {10.1063/1.3076675},
archivePrefix = {arXiv},
       eprint = {0807.2754},
 primaryClass = {astro-ph},
       adsurl = {https://ui.adsabs.harvard.edu/abs/2008AIPC.1085..336Z}
}

@ARTICLE{amato2005,
       author = {{Amato}, E. and {Blasi}, P.},
        title = "{A general solution to non-linear particle acceleration at non-relativistic shock waves}",
      journal = {\mnras},
         year = 2005,
        month = nov,
       volume = {364},
       number = {1},
        pages = {L76-L80},
          doi = {10.1111/j.1745-3933.2005.00110.x},
archivePrefix = {arXiv},
       eprint = {astro-ph/0509673},
 primaryClass = {astro-ph},
       adsurl = {https://ui.adsabs.harvard.edu/abs/2005MNRAS.364L..76A}
}

@ARTICLE{amato2006,
       author = {{Amato}, E. and {Blasi}, P.},
        title = "{Non-linear particle acceleration at non-relativistic shock waves in the presence of self-generated turbulence}",
      journal = {\mnras},
         year = 2006,
        month = sep,
       volume = {371},
       number = {3},
        pages = {1251-1258},
          doi = {10.1111/j.1365-2966.2006.10739.x},
archivePrefix = {arXiv},
       eprint = {astro-ph/0606592},
 primaryClass = {astro-ph},
       adsurl = {https://ui.adsabs.harvard.edu/abs/2006MNRAS.371.1251A}
}

@ARTICLE{blasi2004,
       author = {{Blasi}, Pasquale},
        title = "{Nonlinear shock acceleration in the presence of seed particles}",
      journal = {Astroparticle Physics},
         year = 2004,
        month = apr,
       volume = {21},
       number = {1},
        pages = {45-57},
          doi = {10.1016/j.astropartphys.2003.10.008},
archivePrefix = {arXiv},
       eprint = {astro-ph/0310507},
 primaryClass = {astro-ph},
       adsurl = {https://ui.adsabs.harvard.edu/abs/2004APh....21...45B}
}

@ARTICLE{blasi2002,
       author = {{Blasi}, Pasquale},
        title = "{A semi-analytical approach to non-linear shock acceleration}",
      journal = {Astroparticle Physics},
         year = 2002,
        month = feb,
       volume = {16},
       number = {4},
        pages = {429-439},
          doi = {10.1016/S0927-6505(01)00127-X},
archivePrefix = {arXiv},
       eprint = {astro-ph/0104064},
 primaryClass = {astro-ph},
       adsurl = {https://ui.adsabs.harvard.edu/abs/2002APh....16..429B}
}

@ARTICLE{guo2013,
       author = {{Guo}, Fan and {Giacalone}, Joe},
        title = "{The Acceleration of Thermal Protons at Parallel Collisionless Shocks: Three-dimensional Hybrid Simulations}",
      journal = {\apj},
         year = 2013,
        month = aug,
       volume = {773},
       number = {2},
          eid = {158},
        pages = {158},
          doi = {10.1088/0004-637X/773/2/158},
archivePrefix = {arXiv},
       eprint = {1303.5174},
 primaryClass = {astro-ph.HE},
       adsurl = {https://ui.adsabs.harvard.edu/abs/2013ApJ...773..158G}
}

@article{guo2014b,
doi = {10.1088/0004-637X/797/1/47},
url = {https://dx.doi.org/10.1088/0004-637X/797/1/47},
year = {2014},
month = {nov},
publisher = {The American Astronomical Society},
volume = {797},
number = {1},
pages = {47},
author = {Guo, Xinyi and Sironi, Lorenzo and Narayan, Ramesh},
title = {NON-THERMAL ELECTRON ACCELERATION IN LOW MACH NUMBER COLLISIONLESS SHOCKS. II. FIREHOSE-MEDIATED FERMI ACCELERATION AND ITS DEPENDENCE ON PRE-SHOCK CONDITIONS},
journal = {The Astrophysical Journal}
}

@ARTICLE{wilson2013,
       author = {{Wilson}, III, L.~B. and {Koval}, A. and {Szabo}, A. and {Stevens}, M.~L. and {Kasper}, J.~C. and {Cattell}, C.~A. and {Krasnoselskikh}, V.~V.},
        title = "{Revisiting the structure of low-Mach number, low-beta, quasi-perpendicular shocks}",
      journal = {Journal of Geophysical Research (Space Physics)},
         year = 2017,
        month = sep,
       volume = {122},
       number = {9},
        pages = {9115-9133},
          doi = {10.1002/2017JA024352},
       adsurl = {https://ui.adsabs.harvard.edu/abs/2017JGRA..122.9115W}
}

@ARTICLE{wittor2017,
       author = {{Wittor}, D. and {Vazza}, F. and {Br{\"u}ggen}, M.},
        title = "{Testing cosmic ray acceleration with radio relics: a high-resolution study using MHD and tracers}",
      journal = {\mnras},
         year = 2017,
        month = feb,
       volume = {464},
       number = {4},
        pages = {4448-4462},
          doi = {10.1093/mnras/stw2631},
archivePrefix = {arXiv},
       eprint = {1610.05305},
 primaryClass = {astro-ph.HE},
       adsurl = {https://ui.adsabs.harvard.edu/abs/2017MNRAS.464.4448W}
}

@ARTICLE{diesing2024,
       author = {{Diesing}, Rebecca and {Guo}, Minghao and {Kim}, Chang-Goo and {Stone}, James and {Caprioli}, Damiano},
        title = "{Nonthermal Signatures of Radiative Supernova Remnants}",
      journal = {\apj},
         year = 2024,
        month = oct,
       volume = {974},
       number = {2},
          eid = {201},
        pages = {201},
          doi = {10.3847/1538-4357/ad74f0},
archivePrefix = {arXiv},
       eprint = {2404.15396},
 primaryClass = {astro-ph.HE},
       adsurl = {https://ui.adsabs.harvard.edu/abs/2024ApJ...974..201D}
}

@ARTICLE{diesing2025,
       author = {{Diesing}, Rebecca and {Gupta}, Siddhartha},
        title = "{Nonthermal Signatures of Radiative Supernova Remnants. II. The Impact of Cosmic Rays and Magnetic Fields}",
      journal = {\apj},
         year = 2025,
        month = feb,
       volume = {980},
       number = {2},
          eid = {167},
        pages = {167},
          doi = {10.3847/1538-4357/ada93d},
archivePrefix = {arXiv},
       eprint = {2411.18679},
 primaryClass = {astro-ph.HE},
       adsurl = {https://ui.adsabs.harvard.edu/abs/2025ApJ...980..167D}
}

@ARTICLE{phan2025,
       author = {{Phan}, Vo Hong Minh and {Cristofari}, Pierre and {Peretti}, Enrico and {Tatischeff}, Vincent and {Ciardi}, Andrea},
        title = "{Transient gamma rays from the 2021 outburst of the recurrent nova RS Ophiuchi: the effect of gamma-ray absorption}",
      journal = {arXiv e-prints},
         year = 2025,
        month = apr,
          eid = {arXiv:2504.02043},
        pages = {arXiv:2504.02043},
archivePrefix = {arXiv},
       eprint = {2504.02043},
 primaryClass = {astro-ph.HE},
       adsurl = {https://ui.adsabs.harvard.edu/abs/2025arXiv250402043P}
}

@ARTICLE{strom1994,
       author = {{Strom}, R.~G.},
        title = "{``Guest stars'', sample of completeness and the local supernova rate.}",
      journal = {\aap},
         year = 1994,
        month = aug,
       volume = {288},
        pages = {L1-L4},
       adsurl = {https://ui.adsabs.harvard.edu/abs/1994A&A...288L...1S}
}

@ARTICLE{HESSSNR2018,
       author = {{H.~E.~S.~S. Collaboration} and {Abdalla}, H. and {Abramowski}, A. and {Aharonian}, F. and {Ait Benkhali}, F. and {Ang{\"u}ner}, E.~O. and {Arakawa}, M. and {Arrieta}, M. and {Aubert}, P. and {Backes}, M. and {Balzer}, A. and {Barnard}, M. and {Becherini}, Y. and {Becker Tjus}, J. and {Berge}, D. and {Bernhard}, S. and {Bernl{\"o}hr}, K. and {Blackwell}, R. and {B{\"o}ttcher}, M. and {Boisson}, C. and {Bolmont}, J. and {Bonnefoy}, S. and {Bordas}, P. and {Bregeon}, J. and {Brun}, F. and {Brun}, P. and {Bryan}, M. and {B{\"u}chele}, M. and {Bulik}, T. and {Capasso}, M. and {Caroff}, S. and {Carosi}, A. and {Casanova}, S. and {Cerruti}, M. and {Chakraborty}, N. and {Chaves}, R.~C.~G. and {Chen}, A. and {Chevalier}, J. and {Colafrancesco}, S. and {Condon}, B. and {Conrad}, J. and {Davids}, I.~D. and {Decock}, J. and {Deil}, C. and {Devin}, J. and {deWilt}, P. and {Dirson}, L. and {Djannati-Ata{\"\i}}, A. and {Donath}, A. and {Drury}, L.~O. 'C. and {Dutson}, K. and {Dyks}, J. and {Edwards}, T. and {Egberts}, K. and {Emery}, G. and {Ernenwein}, J. -P. and {Eschbach}, S. and {Farnier}, C. and {Fegan}, S. and {Fernandes}, M.~V. and {Fernandez}, D. and {Fiasson}, A. and {Fontaine}, G. and {Funk}, S. and {F{\"u}{\ss}ling}, M. and {Gabici}, S. and {Gallant}, Y.~A. and {Garrigoux}, T. and {Gat{\'e}}, F. and {Giavitto}, G. and {Giebels}, B. and {Glawion}, D. and {Glicenstein}, J.~F. and {Gottschall}, D. and {Grondin}, M. -H. and {Hahn}, J. and {Haupt}, M. and {Hawkes}, J. and {Heinzelmann}, G. and {Henri}, G. and {Hermann}, G. and {Hinton}, J.~A. and {Hofmann}, W. and {Hoischen}, C. and {Holch}, T.~L. and {Holler}, M. and {Horns}, D. and {Ivascenko}, A. and {Iwasaki}, H. and {Jacholkowska}, A. and {Jamrozy}, M. and {Jankowsky}, D. and {Jankowsky}, F. and {Jingo}, M. and {Jouvin}, L. and {Jung-Richardt}, I. and {Kastendieck}, M.~A. and {Katarzy{\'n}ski}, K. and {Katsuragawa}, M. and {Katz}, U. and {Kerszberg}, D. and {Khangulyan}, D. and {Kh{\'e}lifi}, B. and {King}, J. and {Klepser}, S. and {Klochkov}, D. and {Klu{\'z}niak}, W. and {Komin}, Nu. and {Kosack}, K. and {Krakau}, S. and {Kraus}, M. and {Kr{\"u}ger}, P.~P. and {Laffon}, H. and {Lamanna}, G. and {Lau}, J. and {Lees}, J. -P. and {Lefaucheur}, J. and {Lemi{\`e}re}, A. and {Lemoine-Goumard}, M. and {Lenain}, J. -P. and {Leser}, E. and {Lohse}, T. and {Lorentz}, M. and {Liu}, R. and {L{\'o}pez-Coto}, R. and {Lypova}, I. and {Malyshev}, D. and {Marandon}, V. and {Marcowith}, A. and {Mariaud}, C. and {Marx}, R. and {Maurin}, G. and {Maxted}, N. and {Mayer}, M. and {Meintjes}, P.~J. and {Meyer}, M. and {Mitchell}, A.~M.~W. and {Moderski}, R. and {Mohamed}, M. and {Mohrmann}, L. and {Mor{\r{a}}}, K. and {Moulin}, E. and {Murach}, T. and {Nakashima}, S. and {de Naurois}, M. and {Ndiyavala}, H. and {Niederwanger}, F. and {Niemiec}, J. and {Oakes}, L. and {O'Brien}, P. and {Odaka}, H. and {Ohm}, S. and {Ostrowski}, M. and {Oya}, I. and {Padovani}, M. and {Panter}, M. and {Parsons}, R.~D. and {Pekeur}, N.~W. and {Pelletier}, G. and {Perennes}, C. and {Petrucci}, P. -O. and {Peyaud}, B. and {Piel}, Q. and {Pita}, S. and {Poireau}, V. and {Poon}, H. and {Prokhorov}, D. and {Prokoph}, H. and {P{\"u}hlhofer}, G. and {Punch}, M. and {Quirrenbach}, A. and {Raab}, S. and {Rauth}, R. and {Reimer}, A. and {Reimer}, O. and {Renaud}, M. and {de los Reyes}, R. and {Rieger}, F. and {Rinchiuso}, L. and {Romoli}, C. and {Rowell}, G. and {Rudak}, B. and {Rulten}, C.~B. and {Safi-Harb}, S. and {Sahakian}, V. and {Saito}, S. and {Sanchez}, D.~A. and {Santangelo}, A. and {Sasaki}, M. and {Schlickeiser}, R. and {Sch{\"u}ssler}, F. and {Schulz}, A. and {Schwanke}, U. and {Schwemmer}, S. and {Seglar-Arroyo}, M. and {Settimo}, M. and {Seyffert}, A.~S. and {Shafi}, N. and {Shilon}, I. and {Shiningayamwe}, K.},
        title = "{Population study of Galactic supernova remnants at very high {\ensuremath{\gamma}}-ray energies with H.E.S.S.}",
      journal = {\aap},
         year = 2018,
        month = apr,
       volume = {612},
          eid = {A3},
        pages = {A3},
          doi = {10.1051/0004-6361/201732125},
archivePrefix = {arXiv},
       eprint = {1802.05172},
 primaryClass = {astro-ph.HE},
       adsurl = {https://ui.adsabs.harvard.edu/abs/2018A&A...612A...3H}
}

@ARTICLE{acero2016,
       author = {{Acero}, F. and {Ackermann}, M. and {Ajello}, M. and {Albert}, A. and {Baldini}, L. and {Ballet}, J. and {Barbiellini}, G. and {Bastieri}, D. and {Bellazzini}, R. and {Bissaldi}, E. and {Bloom}, E.~D. and {Bonino}, R. and {Bottacini}, E. and {Brandt}, T.~J. and {Bregeon}, J. and {Bruel}, P. and {Buehler}, R. and {Buson}, S. and {Caliandro}, G.~A. and {Cameron}, R.~A. and {Caragiulo}, M. and {Caraveo}, P.~A. and {Casandjian}, J.~M. and {Cavazzuti}, E. and {Cecchi}, C. and {Charles}, E. and {Chekhtman}, A. and {Chiang}, J. and {Chiaro}, G. and {Ciprini}, S. and {Claus}, R. and {Cohen-Tanugi}, J. and {Conrad}, J. and {Cuoco}, A. and {Cutini}, S. and {D'Ammando}, F. and {de Angelis}, A. and {de Palma}, F. and {Desiante}, R. and {Digel}, S.~W. and {Di Venere}, L. and {Drell}, P.~S. and {Favuzzi}, C. and {Fegan}, S.~J. and {Ferrara}, E.~C. and {Focke}, W.~B. and {Franckowiak}, A. and {Funk}, S. and {Fusco}, P. and {Gargano}, F. and {Gasparrini}, D. and {Giglietto}, N. and {Giordano}, F. and {Giroletti}, M. and {Glanzman}, T. and {Godfrey}, G. and {Grenier}, I.~A. and {Guiriec}, S. and {Hadasch}, D. and {Harding}, A.~K. and {Hayashi}, K. and {Hays}, E. and {Hewitt}, J.~W. and {Hill}, A.~B. and {Horan}, D. and {Hou}, X. and {Jogler}, T. and {J{\'o}hannesson}, G. and {Kamae}, T. and {Kuss}, M. and {Landriu}, D. and {Larsson}, S. and {Latronico}, L. and {Li}, J. and {Li}, L. and {Longo}, F. and {Loparco}, F. and {Lovellette}, M.~N. and {Lubrano}, P. and {Maldera}, S. and {Malyshev}, D. and {Manfreda}, A. and {Martin}, P. and {Mayer}, M. and {Mazziotta}, M.~N. and {McEnery}, J.~E. and {Michelson}, P.~F. and {Mirabal}, N. and {Mizuno}, T. and {Monzani}, M.~E. and {Morselli}, A. and {Nuss}, E. and {Ohsugi}, T. and {Omodei}, N. and {Orienti}, M. and {Orlando}, E. and {Ormes}, J.~F. and {Paneque}, D. and {Pesce-Rollins}, M. and {Piron}, F. and {Pivato}, G. and {Rain{\`o}}, S. and {Rando}, R. and {Razzano}, M. and {Razzaque}, S. and {Reimer}, A. and {Reimer}, O. and {Remy}, Q. and {Renault}, N. and {S{\'a}nchez-Conde}, M. and {Schaal}, M. and {Schulz}, A. and {Sgr{\`o}}, C. and {Siskind}, E.~J. and {Spada}, F. and {Spandre}, G. and {Spinelli}, P. and {Strong}, A.~W. and {Suson}, D.~J. and {Tajima}, H. and {Takahashi}, H. and {Thayer}, J.~B. and {Thompson}, D.~J. and {Tibaldo}, L. and {Tinivella}, M. and {Torres}, D.~F. and {Tosti}, G. and {Troja}, E. and {Vianello}, G. and {Werner}, M. and {Wood}, K.~S. and {Wood}, M. and {Zaharijas}, G. and {Zimmer}, S.},
        title = "{Development of the Model of Galactic Interstellar Emission for Standard Point-source Analysis of Fermi Large Area Telescope Data}",
      journal = {\apjs},
         year = 2016,
        month = apr,
       volume = {223},
       number = {2},
          eid = {26},
        pages = {26},
          doi = {10.3847/0067-0049/223/2/26},
archivePrefix = {arXiv},
       eprint = {1602.07246},
 primaryClass = {astro-ph.HE},
       adsurl = {https://ui.adsabs.harvard.edu/abs/2016ApJS..223...26A}
}

@ARTICLE{Carretero-Castrillo2025,
       author = {{Carretero-Castrillo}, M. and {Benaglia}, P. and {Paredes}, J.~M. and {Rib{\'o}}, M.},
        title = "{New stellar bow shocks and bubbles found around runaway stars}",
      journal = {\aap},
         year = 2025,
        month = feb,
       volume = {694},
          eid = {A250},
        pages = {A250},
          doi = {10.1051/0004-6361/202451336},
archivePrefix = {arXiv},
       eprint = {2502.02658},
 primaryClass = {astro-ph.SR},
       adsurl = {https://ui.adsabs.harvard.edu/abs/2025A&A...694A.250C}
}

@ARTICLE{matsakos2013,
       author = {{Matsakos}, T. and {Chi{\`e}ze}, J. -P. and {Stehl{\'e}}, C. and {Gonz{\'a}lez}, M. and {Ibgui}, L. and {de S{\'a}}, L. and {Lanz}, T. and {Orlando}, S. and {Bonito}, R. and {Argiroffi}, C. and {Reale}, F. and {Peres}, G.},
        title = "{YSO accretion shocks: magnetic, chromospheric or stochastic flow effects can suppress fluctuations of X-ray emission}",
      journal = {\aap},
         year = 2013,
        month = sep,
       volume = {557},
          eid = {A69},
        pages = {A69},
          doi = {10.1051/0004-6361/201321820},
archivePrefix = {arXiv},
       eprint = {1307.5389},
 primaryClass = {astro-ph.SR},
       adsurl = {https://ui.adsabs.harvard.edu/abs/2013A&A...557A..69M}
}

@ARTICLE{marcowith2018,
       author = {{Marcowith}, Alexandre and {Dwarkadas}, Vikram V. and {Renaud}, Matthieu and {Tatischeff}, Vincent and {Giacinti}, Gwenael},
        title = "{Core-collapse supernovae as cosmic ray sources}",
      journal = {\mnras},
         year = 2018,
        month = oct,
       volume = {479},
       number = {4},
        pages = {4470-4485},
          doi = {10.1093/mnras/sty1743},
archivePrefix = {arXiv},
       eprint = {1806.09700},
 primaryClass = {astro-ph.HE},
       adsurl = {https://ui.adsabs.harvard.edu/abs/2018MNRAS.479.4470M}
}

@ARTICLE{tammann1994,
       author = {{Tammann}, G.~A. and {Loeffler}, W. and {Schroeder}, A.},
        title = "{The Galactic Supernova Rate}",
      journal = {\apjs},
         year = 1994,
        month = jun,
       volume = {92},
        pages = {487},
          doi = {10.1086/192002},
       adsurl = {https://ui.adsabs.harvard.edu/abs/1994ApJS...92..487T}
}

@ARTICLE{smartt2009,
       author = {{Smartt}, Stephen J.},
        title = "{Progenitors of Core-Collapse Supernovae}",
      journal = {\araa},
         year = 2009,
        month = sep,
       volume = {47},
       number = {1},
        pages = {63-106},
          doi = {10.1146/annurev-astro-082708-101737},
archivePrefix = {arXiv},
       eprint = {0908.0700},
 primaryClass = {astro-ph.SR},
       adsurl = {https://ui.adsabs.harvard.edu/abs/2009ARA&A..47...63S}
}

@ARTICLE{green2025,
       author = {{Green}, D.~A.},
        title = "{An updated catalogue of 310 Galactic supernova remnants and their statistical properties}",
      journal = {Journal of Astrophysics and Astronomy},
         year = 2025,
        month = jan,
       volume = {46},
       number = {1},
          eid = {14},
        pages = {14},
          doi = {10.1007/s12036-024-10038-4},
archivePrefix = {arXiv},
       eprint = {2411.03367},
 primaryClass = {astro-ph.GA},
       adsurl = {https://ui.adsabs.harvard.edu/abs/2025JApA...46...14G}
}

@ARTICLE{cristofari2020,
       author = {{Cristofari}, Pierre and {Blasi}, Pasquale and {Amato}, Elena},
        title = "{The low rate of Galactic pevatrons}",
      journal = {Astroparticle Physics},
         year = 2020,
        month = dec,
       volume = {123},
          eid = {102492},
        pages = {102492},
          doi = {10.1016/j.astropartphys.2020.102492},
archivePrefix = {arXiv},
       eprint = {2007.04294},
 primaryClass = {astro-ph.HE},
       adsurl = {https://ui.adsabs.harvard.edu/abs/2020APh...12302492C}
}

@ARTICLE{adams2013,
       author = {{Adams}, Scott M. and {Kochanek}, C.~S. and {Beacom}, John F. and {Vagins}, Mark R. and {Stanek}, K.~Z.},
        title = "{Observing the Next Galactic Supernova}",
      journal = {\apj},
         year = 2013,
        month = dec,
       volume = {778},
       number = {2},
          eid = {164},
        pages = {164},
          doi = {10.1088/0004-637X/778/2/164},
archivePrefix = {arXiv},
       eprint = {1306.0559},
 primaryClass = {astro-ph.HE},
       adsurl = {https://ui.adsabs.harvard.edu/abs/2013ApJ...778..164A}
}

@ARTICLE{LHAASO2024,
       author = {{Cao}, Zhen and {Aharonian}, F. and {An}, Q. and {Axikegu} and {Bai}, Y.~X. and {Bao}, Y.~W. and {Bastieri}, D. and {Bi}, X.~J. and {Bi}, Y.~J. and {Cai}, J.~T. and {Cao}, Q. and {Cao}, W.~Y. and {Cao}, Zhe and {Chang}, J. and {Chang}, J.~F. and {Chen}, A.~M. and {Chen}, E.~S. and {Chen}, Liang and {Chen}, Lin and {Chen}, Long and {Chen}, M.~J. and {Chen}, M.~L. and {Chen}, Q.~H. and {Chen}, S.~H. and {Chen}, S.~Z. and {Chen}, T.~L. and {Chen}, Y. and {Cheng}, N. and {Cheng}, Y.~D. and {Cui}, M.~Y. and {Cui}, S.~W. and {Cui}, X.~H. and {Cui}, Y.~D. and {Dai}, B.~Z. and {Dai}, H.~L. and {Dai}, Z.~G. and {Danzengluobu} and {Della Volpe}, D. and {Dong}, X.~Q. and {Duan}, K.~K. and {Fan}, J.~H. and {Fan}, Y.~Z. and {Fang}, J. and {Fang}, K. and {Feng}, C.~F. and {Feng}, L. and {Feng}, S.~H. and {Feng}, X.~T. and {Feng}, Y.~L. and {Gabici}, S. and {Gao}, B. and {Gao}, C.~D. and {Gao}, L.~Q. and {Gao}, Q. and {Gao}, W. and {Gao}, W.~K. and {Ge}, M.~M. and {Geng}, L.~S. and {Giacinti}, G. and {Gong}, G.~H. and {Gou}, Q.~B. and {Gu}, M.~H. and {Guo}, F.~L. and {Guo}, X.~L. and {Guo}, Y.~Q. and {Guo}, Y.~Y. and {Han}, Y.~A. and {He}, H.~H. and {He}, H.~N. and {He}, J.~Y. and {He}, X.~B. and {He}, Y. and {Heller}, M. and {Hor}, Y.~K. and {Hou}, B.~W. and {Hou}, C. and {Hou}, X. and {Hu}, H.~B. and {Hu}, Q. and {Hu}, S.~C. and {Huang}, D.~H. and {Huang}, T.~Q. and {Huang}, W.~J. and {Huang}, X.~T. and {Huang}, X.~Y. and {Huang}, Y. and {Huang}, Z.~C. and {Ji}, X.~L. and {Jia}, H.~Y. and {Jia}, K. and {Jiang}, K. and {Jiang}, X.~W. and {Jiang}, Z.~J. and {Jin}, M. and {Kang}, M.~M. and {Ke}, T. and {Kuleshov}, D. and {Kurinov}, K. and {Li}, B.~B. and {Li}, Cheng and {Li}, Cong and {Li}, D. and {Li}, F. and {Li}, H.~B. and {Li}, H.~C. and {Li}, H.~Y. and {Li}, J. and {Li}, Jian and {Li}, Jie and {Li}, K. and {Li}, W.~L. and {Li}, W.~L. and {Li}, X.~R. and {Li}, Xin and {Li}, Y.~Z. and {Li}, Zhe and {Li}, Zhuo and {Liang}, E.~W. and {Liang}, Y.~F. and {Lin}, S.~J. and {Liu}, B. and {Liu}, C. and {Liu}, D. and {Liu}, H. and {Liu}, H.~D. and {Liu}, J. and {Liu}, J.~L. and {Liu}, J.~Y. and {Liu}, M.~Y. and {Liu}, R.~Y. and {Liu}, S.~M. and {Liu}, W. and {Liu}, Y. and {Liu}, Y.~N. and {Lu}, R. and {Luo}, Q. and {Lv}, H.~K. and {Ma}, B.~Q. and {Ma}, L.~L. and {Ma}, X.~H. and {Mao}, J.~R. and {Min}, Z. and {Mitthumsiri}, W. and {Mu}, H.~J. and {Nan}, Y.~C. and {Neronov}, A. and {Ou}, Z.~W. and {Pang}, B.~Y. and {Pattarakijwanich}, P. and {Pei}, Z.~Y. and {Qi}, M.~Y. and {Qi}, Y.~Q. and {Qiao}, B.~Q. and {Qin}, J.~J. and {Ruffolo}, D. and {S{\'a}iz}, A. and {Semikoz}, D. and {Shao}, C.~Y. and {Shao}, L. and {Shchegolev}, O. and {Sheng}, X.~D. and {Shu}, F.~W. and {Song}, H.~C. and {Stenkin}, Yu. V. and {Stepanov}, V. and {Su}, Y. and {Sun}, Q.~N. and {Sun}, X.~N. and {Sun}, Z.~B. and {Tam}, P.~H.~T. and {Tang}, Q.~W. and {Tang}, Z.~B. and {Tian}, W.~W. and {Wang}, C. and {Wang}, C.~B. and {Wang}, G.~W. and {Wang}, H.~G. and {Wang}, H.~H. and {Wang}, J.~C. and {Wang}, K. and {Wang}, L.~P. and {Wang}, L.~Y. and {Wang}, P.~H. and {Wang}, R. and {Wang}, W. and {Wang}, X.~G. and {Wang}, X.~Y. and {Wang}, Y. and {Wang}, Y.~D. and {Wang}, Y.~J. and {Wang}, Z.~H. and {Wang}, Z.~X. and {Wang}, Zhen and {Wang}, Zheng and {Wei}, D.~M. and {Wei}, J.~J. and {Wei}, Y.~J. and {Wen}, T. and {Wu}, C.~Y. and {Wu}, H.~R.},
        title = "{The First LHAASO Catalog of Gamma-Ray Sources}",
      journal = {\apjs},
         year = 2024,
        month = mar,
       volume = {271},
       number = {1},
          eid = {25},
        pages = {25},
          doi = {10.3847/1538-4365/acfd29},
archivePrefix = {arXiv},
       eprint = {2305.17030},
 primaryClass = {astro-ph.HE},
       adsurl = {https://ui.adsabs.harvard.edu/abs/2024ApJS..271...25C}
}

@BOOK{CTAscience,
       author = {{Cherenkov Telescope Array Consortium} and {Acharya}, B.~S. and {Agudo}, I. and {Al Samarai}, I. and {Alfaro}, R. and {Alfaro}, J. and {Alispach}, C. and {Alves Batista}, R. and {Amans}, J. -P. and {Amato}, E. and {Ambrosi}, G. and {Antolini}, E. and {Antonelli}, L.~A. and {Aramo}, C. and {Araya}, M. and {Armstrong}, T. and {Arqueros}, F. and {Arrabito}, L. and {Asano}, K. and {Ashley}, M. and {Backes}, M. and {Balazs}, C. and {Balbo}, M. and {Ballester}, O. and {Ballet}, J. and {Bamba}, A. and {Barkov}, M. and {Barres de Almeida}, U. and {Barrio}, J.~A. and {Bastieri}, D. and {Becherini}, Y. and {Belfiore}, A. and {Benbow}, W. and {Berge}, D. and {Bernardini}, E. and {Bernardini}, M.~G. and {Bernardos}, M. and {Bernl{\"o}hr}, K. and {Bertucci}, B. and {Biasuzzi}, B. and {Bigongiari}, C. and {Biland}, A. and {Bissaldi}, E. and {Biteau}, J. and {Blanch}, O. and {Blazek}, J. and {Boisson}, C. and {Bolmont}, J. and {Bonanno}, G. and {Bonardi}, A. and {Bonavolont{\`a}}, C. and {Bonnoli}, G. and {Bosnjak}, Z. and {B{\"o}ttcher}, M. and {Braiding}, C. and {Bregeon}, J. and {Brill}, A. and {Brown}, A.~M. and {Brun}, P. and {Brunetti}, G. and {Buanes}, T. and {Buckley}, J. and {Bugaev}, V. and {B{\"u}hler}, R. and {Bulgarelli}, A. and {Bulik}, T. and {Burton}, M. and {Burtovoi}, A. and {Busetto}, G. and {Canestrari}, R. and {Capalbi}, M. and {Capitanio}, F. and {Caproni}, A. and {Caraveo}, P. and {C{\'a}rdenas}, V. and {Carlile}, C. and {Carosi}, R. and {Carqu{\'\i}n}, E. and {Carr}, J. and {Casanova}, S. and {Cascone}, E. and {Catalani}, F. and {Catalano}, O. and {Cauz}, D. and {Cerruti}, M. and {Chadwick}, P. and {Chaty}, S. and {Chaves}, R.~C.~G. and {Chen}, A. and {Chen}, X. and {Chernyakova}, M. and {Chikawa}, M. and {Christov}, A. and {Chudoba}, J. and {Cie{\'s}lar}, M. and {Coco}, V. and {Colafrancesco}, S. and {Colin}, P. and {Conforti}, V. and {Connaughton}, V. and {Conrad}, J. and {Contreras}, J.~L. and {Cortina}, J. and {Costa}, A. and {Costantini}, H. and {Cotter}, G. and {Covino}, S. and {Crocker}, R. and {Cuadra}, J. and {Cuevas}, O. and {Cumani}, P. and {D'A{\`\i}}, A. and {D'Ammando}, F. and {D'Avanzo}, P. and {D'Urso}, D. and {Daniel}, M. and {Davids}, I. and {Dawson}, B. and {Dazzi}, F. and {De Angelis}, A. and {de C{\'a}ssia dos Anjos}, R. and {De Cesare}, G. and {De Franco}, A. and {de Gouveia Dal Pino}, E.~M. and {de la Calle}, I. and {de los Reyes Lopez}, R. and {De Lotto}, B. and {De Luca}, A. and {De Lucia}, M. and {de Naurois}, M. and {de O{\~n}a Wilhelmi}, E. and {De Palma}, F. and {De Persio}, F. and {de Souza}, V. and {Deil}, C. and {Del Santo}, M. and {Delgado}, C. and {della Volpe}, D. and {Di Girolamo}, T. and {Di Pierro}, F. and {Di Venere}, L. and {D{\'\i}az}, C. and {Dib}, C. and {Diebold}, S. and {Djannati-Ata{\"\i}}, A. and {Dom{\'\i}nguez}, A. and {Dominis Prester}, D. and {Dorner}, D. and {Doro}, M. and {Drass}, H. and {Dravins}, D. and {Dubus}, G. and {Dwarkadas}, V.~V. and {Ebr}, J. and {Eckner}, C. and {Egberts}, K. and {Einecke}, S. and {Ekoume}, T.~R.~N. and {Els{\"a}sser}, D. and {Ernenwein}, J. -P. and {Espinoza}, C. and {Evoli}, C. and {Fairbairn}, M. and {Falceta-Goncalves}, D. and {Falcone}, A. and {Farnier}, C. and {Fasola}, G. and {Fedorova}, E. and {Fegan}, S. and {Fernandez-Alonso}, M. and {Fern{\'a}ndez-Barral}, A. and {Ferrand}, G. and {Fesquet}, M. and {Filipovic}, M. and {Fioretti}, V. and {Fontaine}, G. and {Fornasa}, M. and {Fortson}, L. and {Freixas Coromina}, L. and {Fruck}, C. and {Fujita}, Y. and {Fukazawa}, Y. and {Funk}, S. and {F{\"u}{\ss}ling}, M. and {Gabici}, S. and {Gadola}, A. and {Gallant}, Y. and {Garcia}, B. and {Garcia L{\'o}pez}, R. and {Garczarczyk}, M. and {Gaskins}, J. and {Gasparetto}, T. and {Gaug}, M. and {Gerard}, L. and {Giavitto}, G. and {Giglietto}, N. and {Giommi}, P. and {Giordano}, F. and {Giro}, E. and {Giroletti}, M.},
        title = "{Science with the Cherenkov Telescope Array}",
         year = 2019,
          doi = {10.1142/10986},
       adsurl = {https://ui.adsabs.harvard.edu/abs/2019scta.book.....C}
}

@ARTICLE{achterberg1986,
       author = {{Achterberg}, A. and {Blandford}, R.~D.},
        title = "{Transmission and damping of hydromagnetic waves behind a strong shock front - Implications for cosmic ray acceleration}",
      journal = {\mnras},
         year = 1986,
        month = feb,
       volume = {218},
        pages = {551-575},
          doi = {10.1093/mnras/218.3.551},
       adsurl = {https://ui.adsabs.harvard.edu/abs/1986MNRAS.218..551A}
}

@BOOK{mihalas1981,
       author = {{Mihalas}, D. and {Binney}, J.},
        title = "{Galactic astronomy. Structure and kinematics}",
         year = 1981,
       adsurl = {https://ui.adsabs.harvard.edu/abs/1981gask.book.....M}
}

@ARTICLE{sarkar2021,
       author = {{Sarkar}, Kartick C. and {Gnat}, Orly and {Sternberg}, Amiel},
        title = "{Non-equilibrium ionization and radiative transport in an evolving supernova remnant}",
      journal = {\mnras},
         year = 2021,
        month = jun,
       volume = {504},
       number = {1},
        pages = {583-600},
          doi = {10.1093/mnras/stab582},
archivePrefix = {arXiv},
       eprint = {2010.00477},
 primaryClass = {astro-ph.GA},
       adsurl = {https://ui.adsabs.harvard.edu/abs/2021MNRAS.504..583S}
}

@ARTICLE{sutherland2017,
       author = {{Sutherland}, Ralph S. and {Dopita}, Michael A.},
        title = "{Effects of Preionization in Radiative Shocks. I. Self-consistent Models}",
      journal = {\apjs},
         year = 2017,
        month = apr,
       volume = {229},
       number = {2},
          eid = {34},
        pages = {34},
          doi = {10.3847/1538-4365/aa6541},
archivePrefix = {arXiv},
       eprint = {1702.07453},
 primaryClass = {astro-ph.IM},
       adsurl = {https://ui.adsabs.harvard.edu/abs/2017ApJS..229...34S}
}

@BOOK{draine2011,
       author = {{Draine}, Bruce T.},
        title = "{Physics of the Interstellar and Intergalactic Medium}",
         year = 2011,
       adsurl = {https://ui.adsabs.harvard.edu/abs/2011piim.book.....D}
}

\end{document}